\documentclass[a4paper,12pt]{article}\pdfoutput=1
\usepackage{graphicx, rotating, slashed, color, colortbl}
\ifx\pdfoutput\undefined
\usepackage[dvips,bookmarks]{hyperref}	% This is for arXiv.org
\else
\usepackage{hyperref}	% This is for pdftex
\fi
\hypersetup{colorlinks,bookmarksopen,bookmarksnumbered,citecolor=verdes,
linkcolor=blus,pdfstartview=FitH,urlcolor=rossos}
\def\hhref#1{\href{http://arxiv.org/abs/#1}{#1}} % in bibliography

\newcommand{\fig}[1]{~\ref{fig:#1}}
\definecolor{grigio}{cmyk}{0,0,0,0.1}
\definecolor{rosa}{cmyk}{0,0.1,0.1,0.02}
\definecolor{rosino}{cmyk}{0,0.05,0.05,0.02}
\definecolor{rosas}{cmyk}{0,0.3,0.25,0.05}
\definecolor{celeste}{cmyk}{0.1,0,0,0.02}
\definecolor{giallino}{cmyk}{0,0,0.1,0.02}
\definecolor{rosso}{cmyk}{0,1,1,0.4}
\definecolor{rossos}{cmyk}{0,1,1,0.55}
\definecolor{rossoc}{cmyk}{0,1,1,0.2}
\definecolor{blu}{cmyk}{1,1,0,0.3}
\definecolor{blus}{cmyk}{1,1,0,0.5}
\definecolor{bluc}{cmyk}{1,1,0,0.1}
\definecolor{blucc}{cmyk}{0.7,0.5,0,0}
\definecolor{viola}{cmyk}{0,1,0,0.6}
\definecolor{viola2}{cmyk}{0,1,0.2,0.6}
\definecolor{verde}{cmyk}{0.92,0,0.59,0.25}
\definecolor{verdec}{cmyk}{0.92,0,0.59,0.15}
\definecolor{verdes}{cmyk}{0.92,0,0.59,0.4}
\definecolor{verdino}{cmyk}{0.12,0,0.09,0.02}
\definecolor{giallo}{cmyk}{0,0,1,0}
\definecolor{gialloverde}{cmyk}{0.44,0,0.74,0}

\def\SU{{\rm SU}}

\newcommand{\fb}{\,{\rm fb}}
\newcommand{\TeV}{\,{\rm TeV}}
\newcommand{\GeV}{\,{\rm GeV}}
\newcommand{\pb}{\,{\rm pb}}
\newcommand{\beq}{\begin{equation}}
\newcommand{\eeq}{\end{equation}}
\ifx\pdfoutput\undefined
\usepackage[dvips,bookmarks]{hyperref}	% This is for arXiv.org
\else
\usepackage{hyperref}	% This is for pdftex
\fi
\hypersetup{colorlinks,bookmarksopen,bookmarksnumbered,citecolor=verdes,
linkcolor=blus,pdfstartview=FitH,urlcolor=rossos}
\def\hhref#1{\href{http://arxiv.org/abs/#1}{#1}} % in bibliography

\def\circa#1{\,\raise.3ex\hbox{$#1$\kern-.75em\lower1ex\hbox{$\sim$}}\,}

\font\tenrsfs=rsfs10 at 12pt
\font\sevenrsfs=rsfs7
\font\fiversfs=rsfs5
\newfam\rsfsfam
\textfont\rsfsfam=\tenrsfs
\scriptfont\rsfsfam=\sevenrsfs
\scriptscriptfont\rsfsfam=\fiversfs
\def\mathscr#1{{\fam\rsfsfam\relax#1}}

\usepackage{color}
\definecolor{rosso}{cmyk}{0,1,1,0.4}
\definecolor{rossos}{cmyk}{0,1,1,0.55}
\definecolor{rossoc}{cmyk}{0,1,1,0.2}
\definecolor{blu}{cmyk}{1,1,0,0.3}
\definecolor{blus}{cmyk}{1,1,0,0.6}
\definecolor{bluc}{cmyk}{1,1,0,0.1}
\definecolor{verde}{cmyk}{0.92,0,0.59,0.25}
\definecolor{verdec}{cmyk}{0.92,0,0.59,0.15}
\definecolor{verdes}{cmyk}{0.92,0,0.59,0.4}

\newcommand{\be}{\begin{equation}}
\newcommand{\ee}{\end{equation}}
\newcommand{\br}{\begin{eqnarray}}
\newcommand{\bea}{\begin{eqnarray}}
\newcommand{\eea}{\end{eqnarray}}
\newcommand{\er}{\end{eqnarray}}
\newcommand{\ba}{\begin{array}}
\newcommand{\ea}{\end{array}}
\newcommand{\bi}{\begin{itemize}}
\newcommand{\ei}{\end{itemize}}
\newcommand{\bn}{\begin{enumerate}}
\newcommand{\en}{\end{enumerate}}
\newcommand{\bc}{\begin{center}}
\newcommand{\ec}{\end{center}}

\newcommand{\eq}[1]{~(\ref{eq:#1})}

\makeatletter
\def\art{\@ifnextchar[{\eart}{\oart}}
\def\eart[#1]#2#3#4#5#6{{\rm #2}, {#3 #4} {\rm (#6) #5} [{\em arXiv:\hhref{#1}}]}
\def\hepart[#1]#2{{{\rm #2}, \em arXiv:\hhref{#1}}}
\newcommand{\oart}[5]{{\rm #1}, {#2 #3} {\rm (#5) #4}}

\newcounter{alphaequation}[equation]
\def\thealphaequation{\theequation\hbox to
0.6em{\hfil\alph{alphaequation}\hfil}}
\def\eqnsystem#1{
\def\@eqnnum{{\rm (\thealphaequation)}}
\def\@@eqncr{\let\@tempa\relax \ifcase\@eqcnt \def\@tempa{& & &} \or
 \def\@tempa{& &}\or \def\@tempa{&}\fi\@tempa
 \if@eqnsw\@eqnnum\refstepcounter{alphaequation}\fi
\global\@eqnswtrue\global\@eqcnt=0\cr}
\refstepcounter{equation} \let\@currentlabel\theequation \def\@tempb{#1}
\ifx\@tempb\empty\else\label{#1}\fi
\refstepcounter{alphaequation}
\let\@currentlabel\thealphaequation
\global\@eqnswtrue\global\@eqcnt=0 \tabskip\@centering\let\\=\@eqncr
$$\halign to \displaywidth\bgroup \@eqnsel\hskip\@centering
$\displaystyle\tabskip\z@{##}$&\global\@eqcnt\@ne
\hskip2\arraycolsep\hfil${##}$\hfil& \global\@eqcnt\tw@\hskip2\arraycolsep
$\displaystyle\tabskip\z@{##}$\hfil
\tabskip\@centering&\llap{##}\tabskip\z@\cr}
\def\endeqnsystem{\@@eqncr\egroup$$\global\@ignoretrue} \makeatother

\def\gappeq{\mathrel{\rlap {\raise.5ex\hbox{$>$}}
{\lower.5ex\hbox{$\sim$}}}}

\def\lappeq{\mathrel{\rlap{\raise.5ex\hbox{$<$}}
{\lower.5ex\hbox{$\sim$}}}}

       \def\baselinestretch{1.03}

\begin{document}
IFUP-TH/13-2007\hfill CERN-PH-TH/2009-147
\bigskip
\begin{center}
{\LARGE {\color{rossos}\bf Minimal Matter at\\[5mm] the Large Hadron Collider}} \\
\vspace*{1cm}
{\bf  Eugenio Del Nobile}$^a$,
{\bf  Roberto Franceschini}$^b$,\\  {\bf Duccio Pappadopulo}$^b$ and  {\bf Alessandro Strumia}$^{a,c,d}$
%\vspace{0.3cm}

{\em 
$^a$ Dipartimento di Fisica dell'Universit\`a di Pisa, Italia\\
$^b$ ITPP, EPFL, CH-1015, Lausanne, Switzerland \\
$^c$ CERN, PH-TH, CH-1211, Gen\`eve 23, Suisse\\
$^d$ INFN, sezione di Pisa, Italia\\
}
\bigskip\bigskip

\centerline{\large\bf Abstract}
\begin{quote}\large
We classify all possible new ${\rm U}(1)_Y\otimes\SU(2)_L\otimes\SU(3)_c$ 
multiplets that can couple to pairs of SM particles.
Assuming that production of such new particles
is dominated by their gauge interactions we study their signals at LHC, finding the following
five main classes: i) lepto-quark $2\ell \,2q$ signals; ii) di-lepton $4\ell$ signals;
iii) di-quarks $4j$ signals, iv) heavy-lepton $2\ell\, 2V$ signals and v) heavy quarks
$2j\,2V$ signals, where $V$ denotes heavy SM vectors
(with $W$ being associated to exotic fermions).
In each case we outline the most promising final states, the SM backgrounds and 
propose the needed searches.

\end{quote}

\end{center}

\bigskip\bigskip

%\tableofcontents

\section{Introduction}

The Higgs mass hierarchy puzzle suggests new physics around the electroweak scale.
It is usually assumed that new particles carry
a new conserved quantum number, such that
the lightest new particle is a stable Dark Matter candidate, and such that
new physics  affects electroweak precision data only at loop level.
The main scenario is supersymmetry with conserved $R$-parity, where
all new particles are odd under a Z$_2$ matter parity.
Little-Higgs with $T$-parity could be a possible alternative~\cite{Tparity}.
Some authors also consider extra dimensions  with a Z$_2$ orbifold symmetry~\cite{Caccia}.
These scenarios introduce a lot of new particles with unknown masses and
consequently suggest a
huge variety of possible manifestations at LHC.
Their common feature relevant for LHC signatures
is the lack of Z$_2$-odd couplings of the form
\beq\label{eq:Z2odd}\lambda~\hbox{(SM particle) $\cdot$ (SM particle) $\cdot$ (new particle)}.\eeq

%\footnote{We list the main experimental anomalies that could be explained
%in terms of supersymmetry.
%In 2008 the PAMELA and ATIC excesses.
%In 2007 $B_s$ mixing.
%In 2006 the EGRET $\gamma$ excess.
%In 2005, the HyperCP $\Sigma^+\to pX$.
%In 2004 the recompiled {\sc Tristan} $e^+e^-\to {\rm hadrons}$ excess.
%In 2003 $B_S\to \phi K_S$.
%In 2002 the $\ell\gamma \not\!\! E_T$ CDF events.
%In 2001 the $g-2$ of the muon.
%In 2000 the NuTeV long lived neutral state.
%In 1999 $\epsilon'$,
%In 1998
%In 1997 the {\sc Hera} high $Q^2$ excess.
%In 1996 the {\sc Aleph} $4j$ events.
%In 1995 the HEAT excess and the {\sc Karmen} time anomaly.
%Previously, various anomalies in precision data
%($R_b$, $R_c$, $A_{LR}$, $A^b_{\rm FB}$).
%In 1985 the UA1 monojets.}

In section~\ref{MM} we consider a different scenario that suggests well-defined signatures:
we add to the SM
one electroweak multiplet with a mass term and
renormalizable couplings to SM particles
of the form of eq.\eq{Z2odd}.
The Lagrangian is restricted only by imposing the SM gauge and Lorentz symmetries,
not by new symmetries.
We find that  28 possible new multiplets can have such couplings:
13 fermions (table~\ref{tab:listf}) plus
15 scalars (listed in table~\ref{tab:lists});
or equivalently 18 colored (shaded in red) plus 10 uncolored (shaded in blue).

We assume that the couplings $\lambda$ are small enough that 
production of new particles 
is dominated by their SM gauge interactions (weak or strong), and $\lambda$ is only relevant for decays
of new particles, discussed in section~\ref{dec}. 
In this limit precision and flavor data are satisfied~\cite{Raidal}, and one obtains well-defined scenarios of new physics, 
allowing us to study their well-defined signals at LHC, that can be computed in terms of $M$,
up to a minor dependence on $\lambda$ and on its flavor structure.
LHC is the main probe because the smallness of $\lambda$ suppresses the width of
these new particles, not their production cross section.
This is unlike the case of new $Z'$ vectors, that gives signals both at
LHC and in precision data (which did not show deviations from the SM)
only if its coupling constant is large enough.

As well known, dedicated searches are usually necessary to discover new phenomena among the huge backgrounds present at hadron colliders.
These 28 new-physics scenarios give rise to
five main classes of signatures with well defined peaks in appropriate invariant-mass variables
that we discuss in five dedicated sections:
 $4q$ signatures (discussed  in section~\ref{4q}),  
 $4\ell$ (section~\ref{4l}),
 $2\ell 2V$ (section~\ref{2l2V}), 
 $2q2V$ (section~\ref{2q2V}), 
 $2\ell 2q$ (section~\ref{2l2q}), where $q$ denotes SM quarks,
$\ell$ denotes SM leptons and $V=\{W^\pm,Z,h\}$.
Section~\ref{con} contains our conclusions.

\begin{table}[p]
$$\begin{array}{|cccccccc|}\hline
\rowcolor[gray]{0.8}
\hbox{Name}&\hbox{spin}&{\rm U}(1)_Y&\SU(2)_L&\SU(3)_{\rm c} & \hbox{$|Q|=|T_3+Y|$} & \hbox{couplings to}&\hbox{type}
\cr \hline\hline
N &  {1 \over 2}  &\phantom{-}0 & 1 &1 & 0 & LH&\hbox{type-I see-saw} \cr 
\rowcolor{celeste} L' & {1 \over 2} & -{1 \over 2} & 2 &1 & 0, 1  & EH^* &\hbox{LH} \cr
\rowcolor{celeste} E' & {1 \over 2}  &\phantom{-}1 & 1 &1 & 1& LH^*  &\hbox{LH}\cr  
 \rowcolor{celeste}N_3 &  {1 \over 2}  &\phantom{-}0 & 3 &1 &0,1&  LH&\hbox{type-III see-saw}\cr 
\rowcolor{celeste} E_3 &  {1 \over 2}  &\phantom{-}1 & 3 &1 &0,1,2& LH^*&\hbox{LH}\cr 
\rowcolor{celeste}L^{3/2} &  {1 \over 2}  &\phantom{-}{3\over 2} & 2 &1 &1,2&\bar EH^*&\hbox{LH} \cr 
\rowcolor{rosa} Q' & {1 \over 2} &\phantom{-} {1\over 6}  & 2 & 3 & 1/3,2/3 & HU,H^*D&\hbox{QH}\cr
\rowcolor{rosa}U' & {1 \over 2} & -{2 \over 3} & 1 & \bar{3} & 2/3& HQ &\hbox{QH}\cr
\rowcolor{rosa}D' & {1 \over 2} & \phantom{-}{1 \over 3}& 1 &\bar{3} & 1/3 & H^*Q&\hbox{QH}\cr \hline
\rowcolor{rosa}U_3 &  {1 \over 2}  & \phantom{-}{2 \over 3} & 3 & 3 &1/3,2/3,5/3 &\bar QH^* &\hbox{QH}\cr
\rowcolor{rosa}D_3 &  {1 \over 2} &\phantom{-} {1 \over 3} & 3 & \bar 3 &1/3,2/3,4/3 &QH^*&\hbox{QH} \cr
\rowcolor{rosa}Q^{5/6} & {1 \over 2} &\phantom{-} {5\over 6}  & 2 & \bar 3 &1/3,4/3&\bar DH^*&\hbox{QH} \cr
\rowcolor{rosa}Q^{7/6} & {1 \over 2} &\phantom{-} {7\over 6}  & 2 & 3 &2/3,5/3&UH^* &\hbox{QH}\cr
\hline\end{array}$$
\caption{\em\label{tab:listf} List of new fermions that can couple to two SM particles.
$L',E',Q',U',D'$ denote new fermion-antifermions with quantum numbers equal to the corresponding unprimed chiral SM fermions.
Colored (uncolored) particles in red (blue). The last four multiplets involve exotic electric charges.
}
$$\begin{array}{|cccccccc|}\hline
\rowcolor[gray]{0.8}
\hbox{Name}&\hbox{spin}&{\rm U}(1)_Y&\SU(2)_L&\SU(3)_{\rm c} & \hbox{$|Q|=|T_3+Y|$} & \hbox{couplings to}&\hbox{type}
\cr \hline\hline
\rowcolor{celeste}H'& 0 & \phantom{-} {1\over 2} & 2 & 1 &0,1 &\bar L\bar E,QU,\bar Q\bar D&\hbox{second Higgs}\\
\rowcolor{celeste}\tilde{E} & 0 & \phantom{-} 1 & 1 & 1 &1 &LL&\hbox{LL}\\
\rowcolor{celeste}\tilde{E}^2& 0 & \phantom{-} 2 & 1 & 1 & 2 &\bar E\bar E&\hbox{LL}\\
\rowcolor{celeste}\tilde{E}_3& 0 &  \phantom{-} 1 & 3 & 1 &0,1,2&LL,H^*H^*&\hbox{type-II see-saw}\\
\rowcolor{rosa}\tilde{Q} & 0 &\phantom{-} {1\over 6}  & 2 & 3 & 1/3,2/3 & LD&\hbox{LQ}\\
\rowcolor{rosa}\tilde{Q}^{7/6} & 0 &\phantom{-} {7\over 6}  & 2 & 3 &2/3,5/3& LU, \bar E\bar  Q&\hbox{LQ}\\
\rowcolor{rosa}\tilde D & 0& \phantom{-}{1 \over 3}& 1 &\bar{3} & 1/3 & LQ,\bar E\bar U,UD,\bar Q\bar Q &\hbox{LQ/QQ}\cr 
\rowcolor{rosa}\tilde D_3 & 0& \phantom{-}{1 \over 3}& 3 &\bar{3} &1/3,2/3,4/3& LQ,\bar Q\bar Q&\hbox{LQ/QQ} \cr  
\rowcolor{rosa}\tilde D_6 & 0& \phantom{-}{1 \over 3}& 1 &6 &1/3&  U D ,\bar Q\bar Q &\hbox{QQ} \cr  
\rowcolor{rosa}\tilde D_{36} & 0& \phantom{-}{1 \over 3}& 3 &6 &1/3,2/3,4/3& \bar Q\bar Q&\hbox{QQ} \cr  
\rowcolor{rosa}\tilde{U} &0& \phantom{-}{2 \over 3} & 1 & \bar{3} &2/3&\bar D\bar D &\hbox{QQ} \cr
\rowcolor{rosa}\tilde{U}_6 &0& \phantom{-}{2 \over 3} & 1 & \bar{6} & 2/3& \bar D\bar D&\hbox{QQ} \cr
\rowcolor{rosa}\tilde{q}^{4/3}&0& \phantom{-}{4 \over 3} & 1 & \bar{3} &4/3&UU,\bar E\bar D&\hbox{QQ} \cr
\rowcolor{rosa}\tilde{q}^{4/3}_6&0& \phantom{-}{4 \over 3} & 1 & 6 &4/3&UU&\hbox{QQ}  \cr
\rowcolor{rosa}H_8 & 0& \phantom{-} {1 \over 2} & 2 &8 &0,1 &QU,\bar Q\bar D &\hbox{QQ}\cr  
 \hline
 \hline\end{array}$$
\caption{\em\label{tab:lists} List of new scalars that can couple to two SM particles.
Colored (uncolored) particles in red (blue).}
\end{table}

\begin{table}[t]
\centerline{
\begin{tabular}{c| cc}
SU(3) representation & dimension $d$ & Casimir $C$ \\ \hline
singlet & 1 & 0\\
triplet & 3 & 1/2\\
sextet & 6 & 5/2 \\
octet & 8 & 3
\end{tabular}}
\caption{\em Color factors $C$ that enter in $d\hat\sigma/d\hat t$,
defined as ${\rm Tr}T^a T^b = C \delta^{ab}$.
\label{tabC}}
\end{table}

\begin{figure}[t]
\begin{center}
\includegraphics{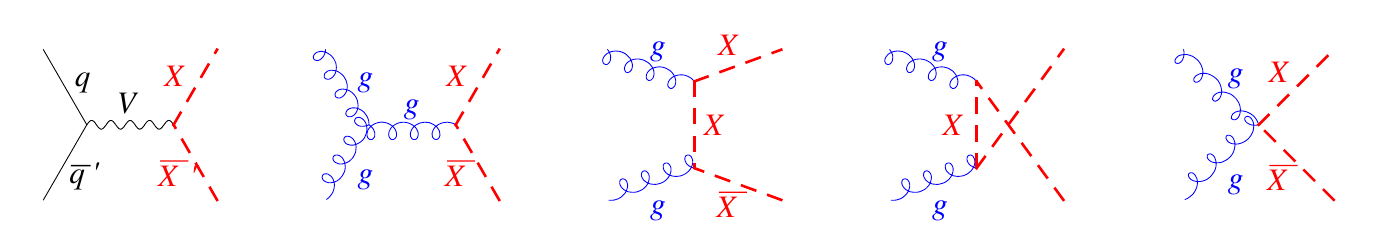}
\caption{\label{fig:Feyn}\em Feynman diagrams for gauge-production of new fermions or scalars $X$
in hadronic collisions.}
\end{center}
\end{figure}

\section{New matter and its production}\label{MM}
We denote the SM fermions as $L,E,Q,U,D$ and the SM Higgs doublet as $H$.
The $L',E',Q',U',D'$ multiplets in table~\ref{tab:listf} 
denote new fermions with the same gauge quantum numbers of the SM ones
plus the corresponding conjugated representations
in order to make  them non-chiral allowing for a gauge-invariant Dirac mass term $M$.
Similarly the first row of table \ref{tab:lists} presents $H'$, i.e.\ a second Higgs doublet.
The other rows list the new exotic particles  
(i.e.\ they have quantum numbers different from SM particles)
that can have cubic couplings with the SM particles.
In order to denote these new particles in a systematic way,
a tilde denote the 15 scalars (so that, in MSSM-like notation, $\tilde{E},\tilde{L},\tilde{Q},\tilde{U},\tilde{D}$ have
the same quantum numbers as the corresponding untilted SM leptons and quarks).
When new particles have $\SU(2)_L$ interactions different than SM particles,
a subscript 3 denotes that they form a triplet under $\SU(2)_L$: for example 
$\tilde{E}_3$ is a scalar triplet with the same couplings as the one that appears in type-II see-saw.
When new particles have non-standard color interactions,
a subscript 6 and 8 denotes they are sextet or octet under color $\SU(3)_c$.
When new multiplet have a  non-standard hypercharge, it is added as superscript.

Up to a few more cubic and quartic scalar couplings involving the Higgs doublet this makes
the full list of possible renormalizable interactions between SM multiplets and one new multiplet.
We do not consider the possibility of adding new massive vectors, because they should be accompanied
by new gauge groups broken by new higgses, giving rise to many new non-minimal possibilities, already studied as extra $Z'$, etc.

\bigskip

The partonic processes that lead to pair production of new particles $X$ (either fermions $\psi$ or scalars $A$)
in $pp$ collisions are
\beq q\bar q\to g, \gamma,Z\to X \bar X \qquad
u\bar d \to W^+ \to X_1 \bar X_2,\qquad
gg\to X \bar X.\eeq
Figure\fig{Feyn} shows the corresponding Feynman diagrams.
The partonic production cross sections, summed over final state colors and polarizations, and averaged over initial state colors and polarizations, are
\begin{eqnsystem}{sys:sigmas} 
\frac{d\sigma}{d\hat t}(q_1 \bar q_2 \to \psi_1 \bar \psi_2)
&=&\frac{V_L^2 + V_R^2}
 {144\pi \hat s^2} 
  (2M_1^2 M_2^2 \!+\! \hat s^2 \!-\! 2(M_1^2+M_2^2)\hat{t}\!+\!2 \hat t^2 \!+\! \hat {s}(2\hat{t}-(M_1\!-\!M_2)^2),\\
\frac{d\sigma}{d\hat t}(q_1 \bar q_2 \to A_1 A^*_2)
&=&\frac{V_L^2 + V_R^2} {144\pi \hat s^2}  (M_1^2 M_2^2  - (M_1^2+M_2^2)\hat{t}+ \hat t^2 + \hat {s}\hat{t} ),\\
\frac{d\sigma}{d\hat t}(gg\to \psi \bar \psi) &=&\frac{ g_3^4C}{8\pi d \hat s^2}
\left[C-\frac{3d}{8\hat s^2}(\hat t -M^2)(\hat u - M^2)\right] f_{g\psi},\\
\frac{d\sigma}{d\hat t}(gg\to AA^*) &=&\frac{g_3^4 C }{8\pi d\hat s^2}\left[C - \frac{3d}{8\hat{s}^2}(\hat t-M^2)(\hat u-M^2)\right]f_{gA}
\end{eqnsystem}
where
\begin{eqnsystem}{sys:f}
f_{g\psi}&=&
\frac{\hat s (4 M^2+\hat s)}{(\hat t-M^2)(\hat u-M^2)}-\frac{4 M^4 \hat s^2}{(\hat t-M^2)^2(\hat u-M^2)^2}-2,
%-\frac{2 M^8-8 t M^6+\left(3 s^2+4 t s+12 t^2\right) M^4-\left(s^3+2 t s^2+8 t^2 s+8 t^3\right) M^2+t (s+t) \left(s^2+2 t s+2 t^2\right)}{ \left(M^2-t\right)^2 \left(-M^2+s+t\right)^2}
  \\
f_{gA}&=&1-\frac{2M^2\hat s}{(\hat t-M^2)(\hat u-M^2)}+\frac{2M^2\hat s^2}{(\hat t-M^2)^2(\hat u-M^2)^2}.
\end{eqnsystem}
$\hat u=M_1^2+M_2^2-\hat s-\hat t$,
%$N_c = 3$, $\beta\equiv \sqrt{1-4M^2/\hat s}$ is the velocity ($0\le\beta\le 1$) of the new particle in the partonic CM frame,
$d$ and $C$ are the color dimension and the Casimir of the new particle as given in table~\ref{tabC}.\footnote{
Given the new particles listed in tables~\ref{tab:listf} and~\ref{tab:lists}, pair production of 
real scalars or fermions is never relevant for us.
In such cases the cross sections must be divided by 2, and with this specification
our foruml\ae{} are fully general, allowing e.g.\ to compute production of an
electro-weak neutral color octet scalar or fermion, such as the supersymmetric Majorana gluino}.
The $q_1\bar q_2\to A_1A_2^*$ amplitude
is $p$-wave suppressed when the scalars $A_{1,2}$ are
non-relativistic, so that scalar production has a lower cross section than production of fermions
with the same gauge charges.
We defined:
\beq\label{eq:V}\begin{array}{ll}
V_A^2= \displaystyle 
8C_q C_X\bigg( \frac{ g_3^2}{\hat s}\bigg)^2+3d
\bigg(Q_q Q_X \frac{ e^2}{\hat s}+
g_q^A g_X\frac{g_2^2/c_{\rm W}^2}{\hat{s} - M_Z^2}\bigg)^2
\qquad& \hbox{for $q\bar q\to X\bar X$ }\\
V_A^2 = 3cd\displaystyle\bigg(\frac{g_2^2}{\hat s-M_W^2}\bigg)^2 & \hbox{for $u\bar d\to X_1X_2$ }
\end{array}\eeq
where $c\neq 0$ only if $A=L$ and $T_3(X_1) - T_3(X_2)=\pm1$:
$c=1$ if $X$ is a weak doublet;
$c=2$ if  $X$ is a weak triplet;
$g_q^A = T_3 - s_W^2 Q_q$ is the $Z$ coupling of quark $q$ with helicity $A=\{L,R\}$.
As the $X$ particles possess an $\SU(2)_L$-invariant mass term $M$, their couplings are vector-like.

\begin{figure}
\begin{center}
\includegraphics[width=0.45\textwidth,height=0.6\textheight]{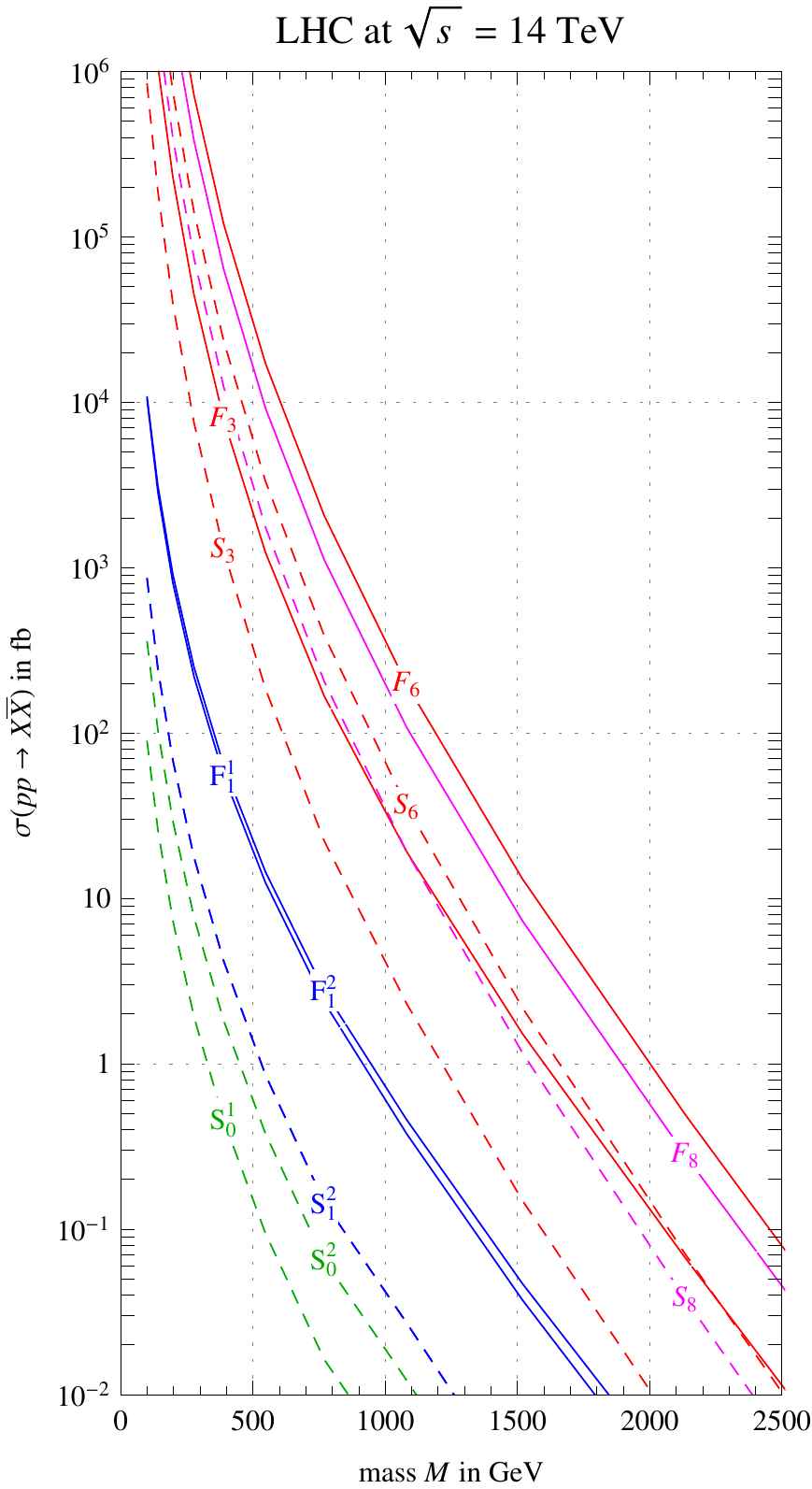}\qquad
\includegraphics[width=0.45\textwidth,height=0.6\textheight]{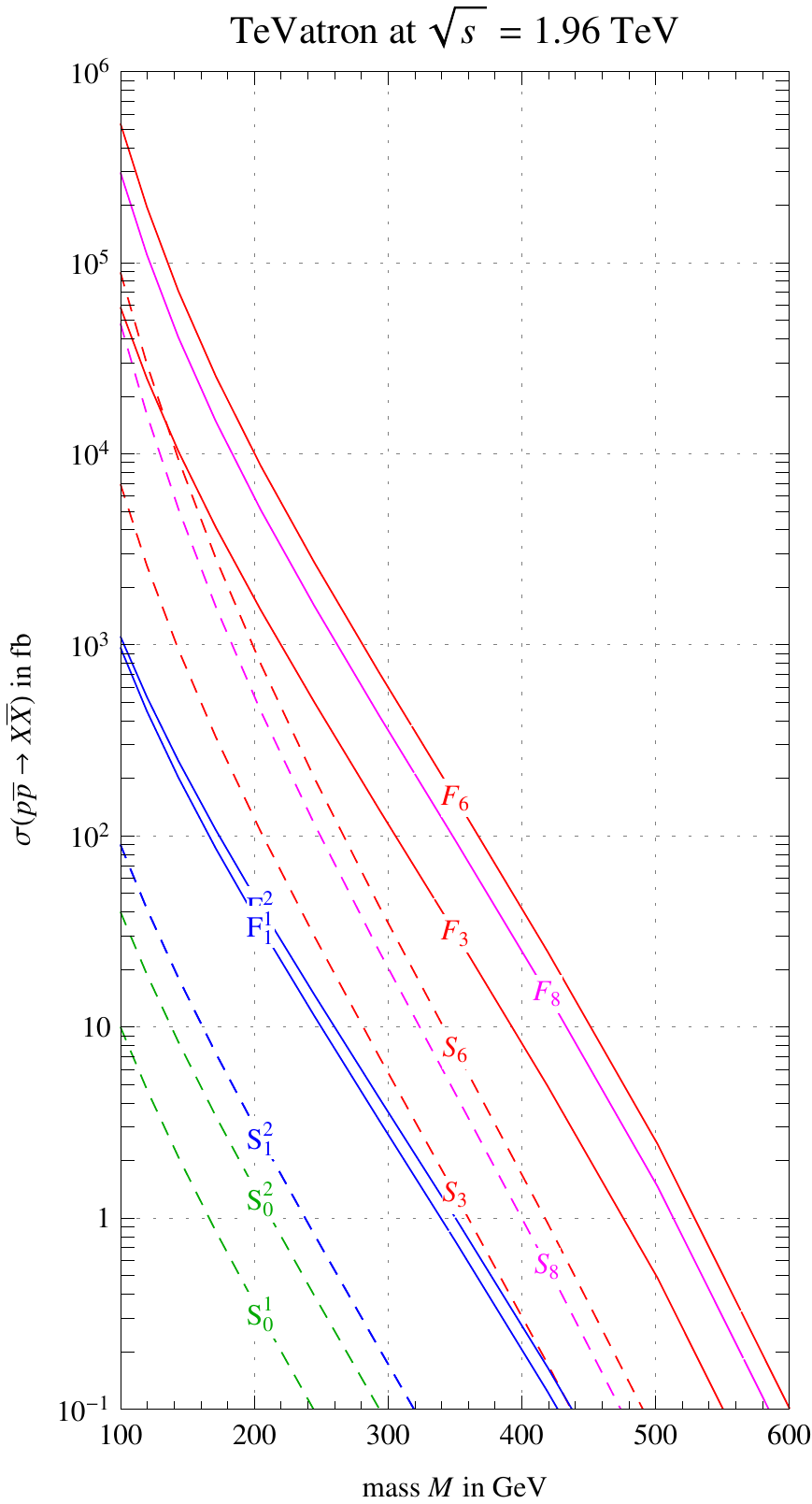}
\caption{\label{fig:sigma}\em Cross sections for pair production via gauge interactions at leading order
in hadronic collisions
of new particles $X\bar X$ labelled as
$P^Q_{T_3}$ where $Q$ is the electric charge,
$T_3$ is weak isospin and
$P=F~(S)$ for a fermion (scalar).
Cross sections of colored particles (in red) negligibly depend on their electroweak interactions,
so we adopted the simplified notation $P_d$ where $d=\{3,6,8\}$
for color triplets, sextets and real color octets.
Couplings and MSTW 2008 pdf are renormalized at $Q^2=\hat{s}$.}
\end{center}
\end{figure}

The resulting proton-proton cross section is
\beq \frac{d\sigma}{dt}(pp\to F) =\sum_{p_1,p_2} \int_0^1 dx_1 dx_2 ~\wp_{p_1}(x_1,Q)\wp_{p_2}(x_2,Q)
 \frac{d\sigma}{d\hat t}(p_1 p_2\to F)\eeq 
where one must sum over all partons $p_{1,2}$ 
with parton distribution functions $\wp_{p_i}(x_{i},Q)$
and we assume $Q^2=\hat{s}$.
We defined $\hat s = s x_1 x_2$, $\hat t = x_1 x_2 t$
and it is convenient to change variables to $X=x_1 x_2$ and $y=\ln(x_1/x_2)/2$
 (i.e.\ $x_{1,2}=\sqrt{X} e^{\pm y}$) such that $dx_1 \,dx_2/x_1x_2 = dy ~dX/X$
 with $|y|<-\ln X/2$.

Fig.\fig{sigma}a  shows the total pair production cross section (no cuts imposed) of a few sample particles
$X\bar X$ as function of their mass $M$ at the LHC $pp$ collider
with planned energy $\sqrt{s}=14\TeV$ and planned luminosity
${\cal L}=300/\,{\rm fb}\cdot{\rm yr}$
as well as planned starting date 2007.\footnote{
Should LHC reach only a fraction $r\sim 1/2$ of its planned energy,
the reduced cross section are roughly obtained modifying the
masses $M$ on the horizontal axis of fig.\fig{sigma}a as $M\to r\cdot M$.}
%Not forgetting that a $\SU(2)_L$ multiplet can contain more particles,
%we labelled the various particles $X$ with the notation
%$$\{ \hbox{color}~ ,~ \hbox{electric charge} ~,~ T_3 ~,~ \hbox{$S$calar or $F$ermion} \}.$$
One can see the expected trends.
The largest possible cross section if for color octet or sextets, similar in the two cases.
The lowest cross section is obtained for particles with only hypercharge interactions.

Fig.\fig{sigma}b shows the corresponding cross sections at the 
 TeVatron $p\bar p$ collider at $\sqrt{s}=1.96\TeV$.
The CDF and D0
experiments published bounds down to $\sigma\circa{<}\hbox{few fb}$ on various processes.

We ignore higher order QCD processes, that lead to pair production of new particles together with jets.
Furthermore, if $\lambda$ is large enough,
at very large masses $M$ the cross section for
single production $X$ via the cubic coupling $\lambda$ (possibly at one loop)
becomes larger than the cross section for pair $X\bar X$ production via SM gauge interactions.
We do not explore this possibility, that leads to different signatures.

\section{Mass spectra and decays}\label{dec}
We first need to compute the mass spectrum of the new particles.
The mass term $M$ gives a common mass $M$ to all components of the new weak multiplet.
If it is a scalar multiplet, it can have a quartic coupling $(X^\dagger T^a X)(H^\dagger  T^a H)$
to the Higgs doublet $H$ that splits the $X$ components
according to their $T^3$.
For both scalars and fermions,  electroweak corrections generate a mass splitting
among the components of $\SU(2)_L$ multiplets.
Specializing eq.~(6) of~\cite{MDM} to the case
$M\gg M_W$,
the mass difference between two components with electric charge $Q$ and $Q+1$ is
\beq  \Delta M=M_{Q+1} - M_Q = (1+2 Q+ \frac{2Y}{\cos\theta_{\rm W}})\alpha_2 M_W \sin^2\frac{\theta_{\rm W}}{2}= 166\,{\rm MeV} (1+2 Q+ \frac{2Y}{\cos\theta_{\rm W}}).
\label{eq:166}\eeq
This means that the lightest component is the one with the smallest electric charge.
(The neutron is instead heavier than the proton because they are composite of quarks with different masses;
we here consider new elementary particles).

\medskip

Therefore two competing effects lead to decays of the new particles.
The $\lambda$ couplings give rise to their decays into SM particles.
Furthermore, heavier components of the multiplets have electroweak decays into the lighter components,
with  rates suppressed by the small phase space.
Since $\Delta M  >m_\pi$  two-body decays into pions are open,
and for both scalars and fermions one has 
\beq
\Gamma(X_{Q+1} \to X_Q \pi^+   ) = 
\displaystyle c
\frac{G_{\rm F}^2V_{ud}^2\, \Delta M^3 f_\pi^2}{\pi}
\sqrt{1-\frac{m_\pi^2}{\Delta M^2}}\sim \frac{1}{\rm mm}\label{gauge_width}\eeq
where $c=2$ for a weak triplet and $c=1$ for a weak doublet.
If $\Delta M\circa{>}\GeV$ one must consider decays into two $\pi$
and ultimately compute the decay at the quark level.

The couplings $\lambda$ generically lead to a decay rate
\beq \Gamma \sim \frac{M\lambda^2}{4\pi}
\sim\frac{1}{3\,{\rm cm}} \frac{M}{\TeV}\frac{\lambda^2}{10^{-16}}. \label{yuka_width}\eeq

%We focus on the decay modes.
%For a new scalar decaying into two light SM fermions (either QQ or LL or QL)
%one simply has
%\beq \Gamma(A\to f_1 \bar f_2) = \frac{\lambda^2}{8\pi}M
%\eeq
%where, in cases where color contractions are involved,
%the coupling $\lambda$ is normalized such that the above formula holds.

%%% Gamma(lambda) > Gamma(Gauge) quando lambda > 5 x 10^-8
%%% uno scambio a tree-level di uno dei nostri scalari produce un operatore di dim=6 con coeff.
%%% lambda^2/M^2 < (1/10^4TeV)^2 per il flavor. da cui lambda < 10^-4 (10^-5)  per M=1 TeV (100 GeV).
%%% la regioen di lambda tra 5x10^-8 e 10^-4 (10^-5) e' ok col flavor e i decay sono dati da lambda (e non da int. di gauge) 

For a new fermion coupled to a SM fermion (either Q or L)
and to the SM Higgs doublet, a generic simple result holds in the $M\gg M_W$
SU(2)$_L$ invariant limit: 
all components of the multiplet have the same decay rate.
Furthermore, when two channels are allowed
(one involving the upper component $H^+$ and the other the lower $H^0$
component of the Higgs doublet),
their relative widths are fixed 
by $\SU(2)_L$ group theory to be 1 or 0, as easily read
from the Yukawa Lagrangian in the gauge-less limit.

Taking into account gauge interactions adds corrections suppressed by $M_W^2/M^2$:
indeed, once that the Higgs gets a vev, $H^0 = v + (h + i\eta)/\sqrt{2}$
and the $V=\{Z,W^\pm\}$ vectors become massive by `eating' the Goldstones
$\eta$ and $H^+$ in the Higgs doublet $H$,
decays into $H^\pm$ are replaced by decays into $W^\pm$,
and decays into $H^0$ by decays into $h,Z$ with equal BR.
Concretely, this arises because inserting the Higgs vev in the Yukawa couplings $FHf$
generates a mass mixing between the new fermions $F$ and the SM fermions $f$,
such that gauge interactions of $F$ and of $f$ become gauge interactions of $F$ with $f$.
Fermions with exotic electric charges cannot mix with SM fermions and
thereby they always lie in weak multiplets together with non-exotic new fermions that can mix.

Notice that experimentalists searched for `excited leptons' or `excited quarks'
that decay into leptons or quarks plus a photon.  Photons are not generated by our 
`heavy leptons' or `heavy quarks', because they couple to the Higgs doublet which is `eaten'
by $Z$ and $W$ but not by photons.
Ignoring photons and the physical Higgs boson $h$ (just because its phenomenology depends on its still unknown mass)
we list the decays  of new heavy particles into pairs of SM particles
allowed by electric charge conservation
and by Lorentz invariance:
$$\begin{array}{c ccc}
\hbox{charge} && \hbox{if scalar} & \hbox{if fermion}\\
0 &\to &f\bar f, W^+W^-,ZZ & \nu Z,~W^\pm \ell^\mp\\
1/3 &\to & \bar u \ell^+,~\bar d\nu,~ud  & \bar d Z,~\bar u W^+\\
2/3 &\to & d\ell^+,~u\nu,~\bar d \bar d & uZ, ~dW^+\\
1 &\to & \ell^+ \nu ,~ u\bar d,~W^+Z& \ell^+ Z, ~\nu W^+ \\
4/3 &\to & uu & \bar d W^+\\
5/3 &\to & u \ell^+ & u W^+\\
2 &\to & \ell^+\ell^+,~W^+W^+ & \ell^+W^+
\end{array}$$
where $f$ denotes any SM fermion.

% involves two channels with equal contributions $\chi^+ f$ and 

%
%decays into one SM fermion and one 
%component of the Higgs doublet $H$
%($\chi^-=h+i \eta$ and $\chi^+$, where
%$h$ is the physical Higgs and $\eta,\chi^\pm$ are the
%Goldstones to be eaten by the massive SM vectors $V=\{Z,W^\pm\}$)

%have equal BR and

%

%
%Decays involve with comparable BR all the four components of the Higgs doublet $H$: 
%the physical higgs $h$ and the Goldstones $\eta,\chi^\pm$, `eaten' by the massive vectors $V=\{Z,W^\pm\}$.
%Decays into $\eta,\chi^\pm$ are replaced by decays into the longitudinal components of $Z,W^\pm$,
%with equal BR up to corrections suppressed by $M_{W,Z}^2/M^2$.
%For a new fermion $\psi$ this happens as follows.
%Inserting the Higgs vev into the Yukawa coupling $\epsilon\,FH\psi$ of $\psi$ to a SM fermion $F$,
%gives rise to a mass mixing between $\psi$ and $F$, such that, rotating to  mass eigenstates, 
%the SM $Z$ gauge interaction $\bar F \slashed{V} F$
%develops a $\bar F \slashed{V} \psi$ coupling proportional to $\epsilon$.
%For a scalar $A$, its mass mixing with the Higgs transforms
%$|D_\mu H|^2 \in H^* H (V+V^2)$ into $ \epsilon A(Vh+V^2 v)$.
%\footnote{A similar mechanism should be operative for
%$ALLH$ case???}

%

%Fermion multiplets couple to the Higgs getting a mass mixing with SM fermions
%and consequently a coupling to SM vectors.

\bigskip

%We can now discuss the phenomenological aspects of the decays.
In the following we will discuss all these possibilities, except the decay of a scalar into a pair of vectors which is not present in our scenario. This kind of decay appears however in other contexts as discussed in \cite{vectorlike}.

When $\lambda$ is such that the gauge decay width eq. (\ref{gauge_width}) is not negligible the decay length could be macroscopically large, leading to the usual associated extra signatures. This is a bonus selection criterion for all the signatures discussed in the following sections.
However the experimental resolution on displaced vertexes is about 100 $\mu m$ which means that for small values of $\lambda$ the displacement is undetectable. As such \emph{detectably} displaced vertex arise only in a small range within the allowed range for $\lambda$ from current experimental limits. 
Furthermore  for decays mediated by gauge interactions the displaced vertex is likely to be undetectable because (at least for new fermions)
$\Delta M$ is so small that the $\pi^\pm$ emitted in the decay are too soft.
%and are in any case experimentally challenging. %Therefore we will not make any use of this potentially interesting feature. 
A possibility to see this kind of decays could be  a dedicated trigger thought to infer
a kink in a charged track to the change in the
electric charge of the heavy particle $X$
that decays in the detector magnetic field.
However if $X$ is colored it hadronizes before decaying, 
forming $X$-hadrons with various possible electric charges
(even changing during its interactions with the detector material)
and washing out the effect~\cite{HadCol}.
%spin correlations between $X$ decay products
In the case where electroweak decays cannot be seen and dominate over $\lambda$ decays,
one effectively has production of the lightest component
of the weak multiplet with cross section equal to production of all multiplet components.
In the following we will focus on the opposite case of prompt $\lambda$ decays, that dominate
over electroweak decays, so that each multiplet gives rise to a set of different signatures.

%We can divide them into five main classes of signals, 
We can divide the signals into five main classes, discussed in the next five sections.
In all cases we consider the standard set of isolation and detection cuts needed to
make the leptons and jets identifiable:
\beq
p_{T}>20 \GeV,\qquad\Delta R>0.4,\qquad |\eta |<2.5 . \label{gencuts}
\label{eq:gencuts}
\eeq
for all particles.
Here $p_T$ is the momentum orthogonal to the beam axis,
$\eta$ is the pseudo-rapidity and $\Delta R=(\Delta\phi^2_T+\Delta\eta^2)^{1/2}$,
with $\Delta \phi_T$ being the angular separation in the plane $T$ransverse to the beam.

All our signals have been computed using a MonteCarlo code written by us in {\sc Mathematica} and using MSTW 2008 PDFs \cite{mstw08}. Some of them have also been computed using {\sc Madgraph}~\cite{MadGraph} where the new particles interactions were added using FeynRules~\cite{FR}. All SM backgrounds are computed using either {\sc MadGraph} or {\sc AlpGen}~\cite{AlpGen} using PDFs CTEQ6L1 and CTEQ5L respectively \cite{cteq6l1}. In the cases where showering has been considered it has been performed with {\sc Pythia 8.1}~\cite{pythia}.

\begin{figure}[t]
\begin{center}
\includegraphics[width=0.45\textwidth,height=55mm]{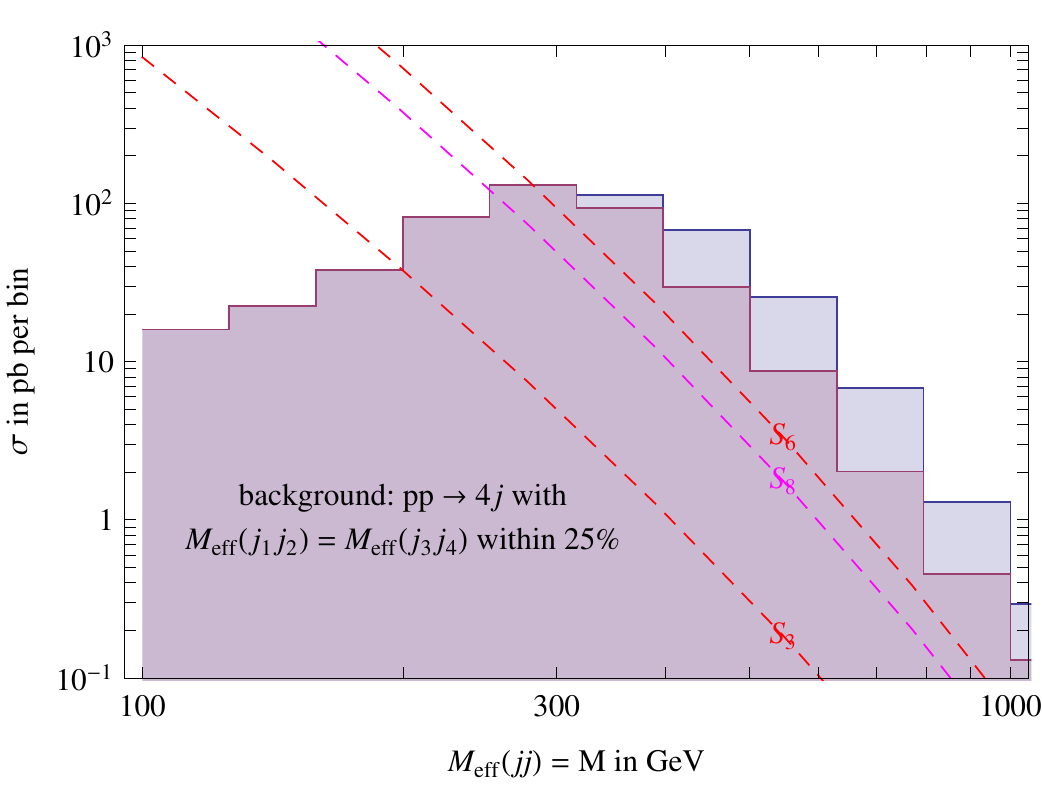}\qquad
\raisebox{-2mm}{\includegraphics[width=0.45\textwidth,height=55mm]{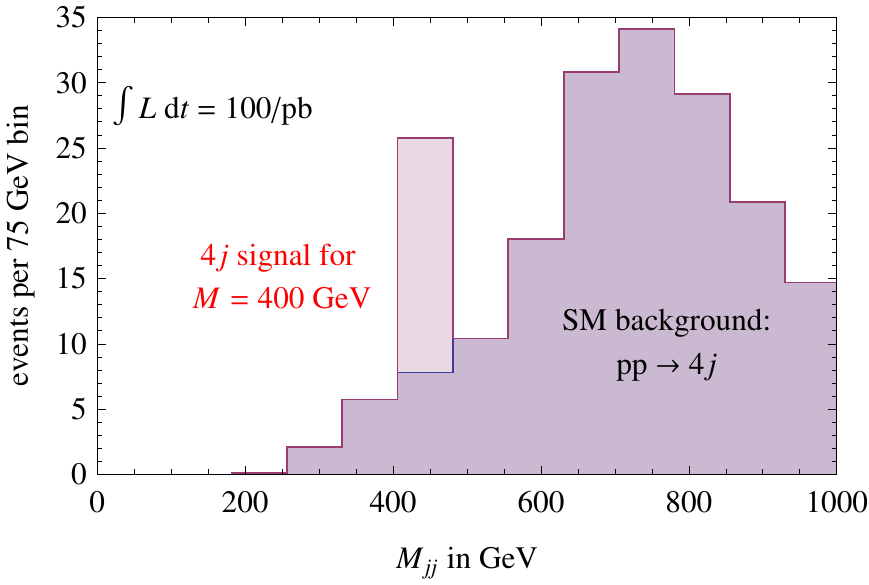}}
\caption{\label{fig:4j}\em cross section $\sigma$ of
$4j$ signals and SM background as function of $M$ (left);
distribution  $d\sigma/dM_{jj}$ for $M=400\GeV$ (right).
}
\end{center}
\end{figure}

\section{$4q$ di-quark signals}\label{4q}
We here consider scalars that couple to two quarks.
One possibility is a second Higgs doublet, which has a complex and well explored phenomenology
that we will not discuss here.
All other possibilities involve colored scalars, that can transform under $\SU(3)_c$ as an octet
 (a `colored higgs'~\cite{colorons}), sextet or triplet: in all cases their production cross sections are large and dominated
 by strong interactions.
The color representation affects the total signal rate 
(triplets have lower production cross sections than octets or sextuplets)
and mildly affects the shape of
$d\hat\sigma/d\hat t$.
 Various weak multiplets are possible, containing electric charges which are either 0 and 1
(second higgs or colored higgs) or various combinations of $1/3$, $2/3$ and $4/3$.
They cannot be discriminated as long as each new particle decays into two light quarks, so that
the only observable signal is $4j$.  The main main background to this signal
are the QCD $4j$ events. We perform the following cuts:
\begin{itemize}
\item[i)] $|\eta|<2.5$ for all jets and $\Delta R>0.4$ for all jet pairs;
\item[ii)] we select among the $4j$ the couple of pairs with ratio $R$ of their effective masses closer to 1,
and accept the event if $R$ deviates from 1 by less than $25\%$, determining
the $M_{\rm eff}$ associated to the event;
\item[iii)] $p_T >\max(r M_{\rm eff},100\GeV)$ for each jet with $r=0.2$ or $r=0.3$;
\item[iv)]  $H_T \equiv \sum_j p_{Tj}>2 M_{\rm eff}$.
\end{itemize}
Fig.\fig{4j}a shows the resulting background as function of $M_{\rm eff}$,
choosing bins of $25\%$ size such that the signal is a peak entirely concentrated in a single bin,
corresponding to the mass $M$ of the new scalar.
The curves show the total signal cross section as function of $M$;
the acceptance of the signal after these cuts is $30\%$ for $r=0.2$ and $12\%$ for $r=0.3$.
Even if the signal is somewhat below the background, 
event rates are so large that
the statistical significance of the signal peak,
$N_{\rm ev}/\sqrt{N_{\rm bck}}$, can allow its detection,
provided that the background rate can be independently computed.
This can be done by relying on the larger sample of QCD $4j$ events
such that $2j$ pairs do not have the same invariant mass.
The signal/background ratio can be enhanced by devising
cuts that optimize the discrimination of the signal from the background:
for example taking into account that QCD jets tend to be forward and hierarchical in $p_T$
while the signal tends to give central jets paired in cones.

\begin{figure}[t]
\begin{center}
{\includegraphics[width=0.45\textwidth,height=55mm]{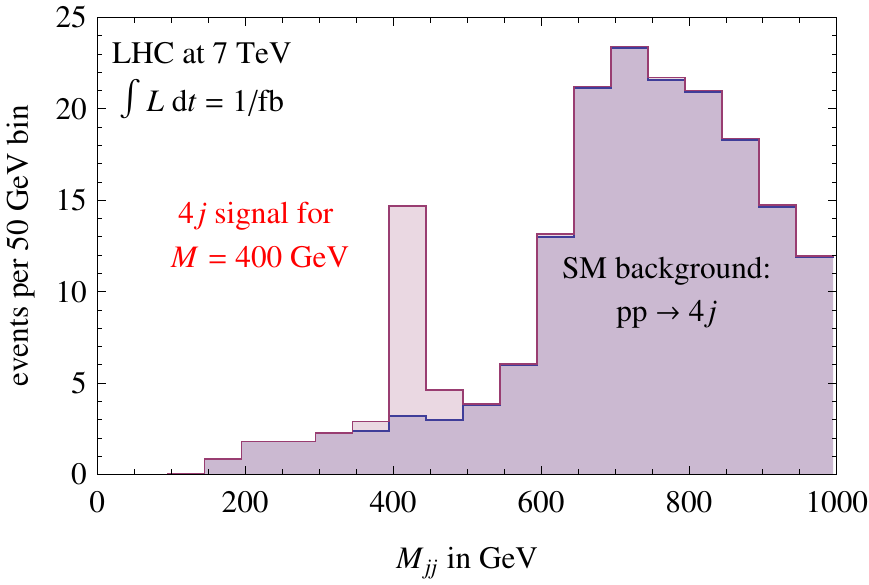}}
\caption{\label{fig:4jlhc7}\em distribution  $d\sigma/dM_{jj}$ for $M=400\GeV$ at the $7 \TeV$ LHC for 1/fb of integrated luminosity.
}
\end{center}
\end{figure}

Although the discovery by event counting described above seems pretty reasonable the existence of new resonances in multi-jet final states can be established in a more robust way looking for a peak in the invariant mass of suitably chosen jet pairs. This would allow to make a discovery without any theoretical input about the background shape and normalization.
To exemplify how to achieve this goal, we consider one scalar octet with mass $M=400\GeV$ such that the signal cross section computed with   {\sc Madgraph} and CTEQ6L1 PDFs  is $\sigma = 8\pb$ after the basic cuts of eq.~(\ref{eq:gencuts}). Besides
performing the cuts described above, we also require
\begin{itemize}
\item[v)] $p_T^j>250\GeV$ and $\Delta \eta_{jj}<1.7$.
\end{itemize}
%In each event we retain only the pairing whit the least difference between jet pair invariant masses. 
The distribution of the retained jet pair invariant mass is shown in fig.\fig{4j}b assuming 100/pb of
integrated luminosity: a clear peak above the background is present at the mass $M$ of the new
particle.  Additionally, in figure \ref{fig:4jlhc7}  we present the result for the same analysis in the case of 1/fb of integrated luminosity for the LHC running at $7 \TeV$. 

We verified that along the same lines, a heavier $M= 1 \TeV$ scalar  gives
a clear peak with an integrated luminosity of 100/fb. 

%	
%	Plotting the invariant mass of the and plot the invariant mass of the closest in fig.\fig{} the background and the signal as function of $M_{jj}$
%	in bins with 75 GeV width for an integrated luminosity of 100/pb. 
%	The signal is the clearly visible peak at $M_{jj}=M$.
%A similar plot can be done for $M=1\TeV$ and 100/fb.

\smallskip

If one or more of the four signal quarks is a top quark, a different and easier signal is obtained.
In our scenario the production cross section is entirely due to QCD gauge interactions.
A somewhat related $4j$ signal was previously studied in~\cite{colorons}, where production can be mediated by a new particle, such that the invariant mass of all $4j$ is around its mass,
and a larger cross section can be obtained.

\section{$4\ell$ di-lepton signals}\label{4l}
We consider scalars that couple to two leptons.
There are four possible complex scalars.
Two of them have already been studied as `type II see-saw`~\cite{typeII}
and `second Higgs doublet': in both cases they can also couple 
to the Higgs or to quarks, so that the name `di-lepton' is not fully appropriate for them.
The pure di-leptons are the remaining two cases: 
\begin{itemize}
\item[i)]
the singlet $\tilde{E}$ with $Y=1$
 and Yukawa coupling $\tilde{E}L_iL_j= \tilde{E} (\ell_i \nu_j - \ell_j\nu_i)$ and
\item[ii)] the singlet $\tilde{E}^2$ with $Y=2$ and Yukawa coupling
$\tilde{E}^2 EE$.
\end{itemize}
They have different signatures.

\begin{figure}
\begin{center}
\includegraphics[width=0.45\textwidth]{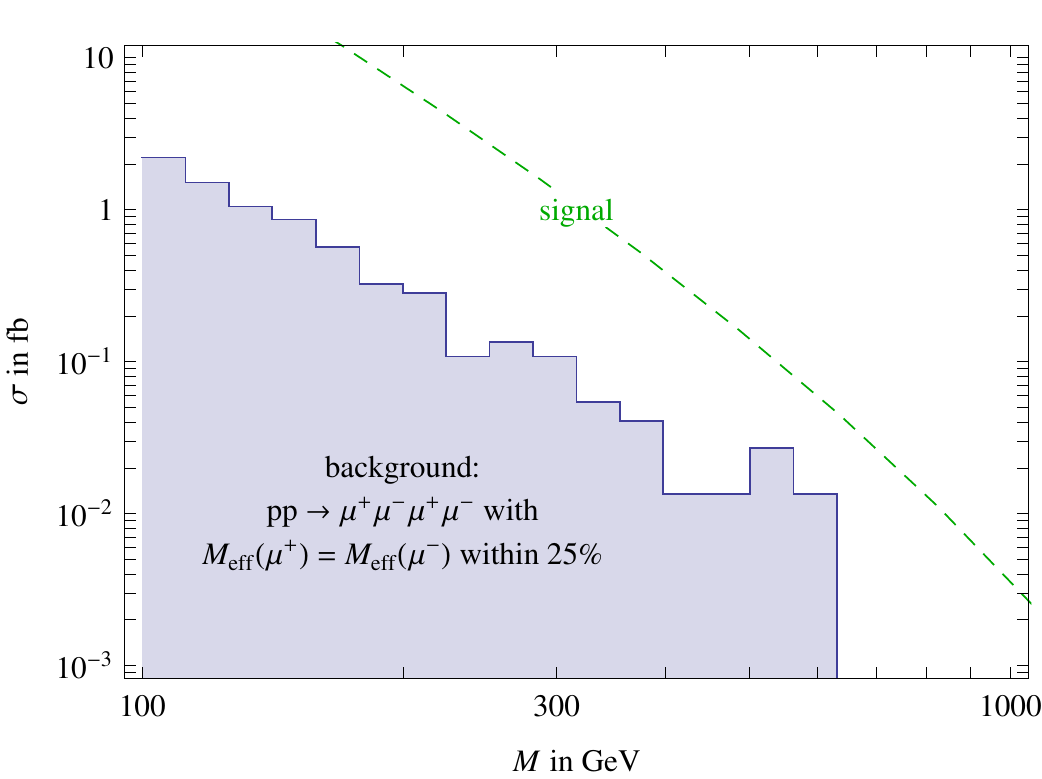}\quad
\includegraphics[width=0.5\textwidth]{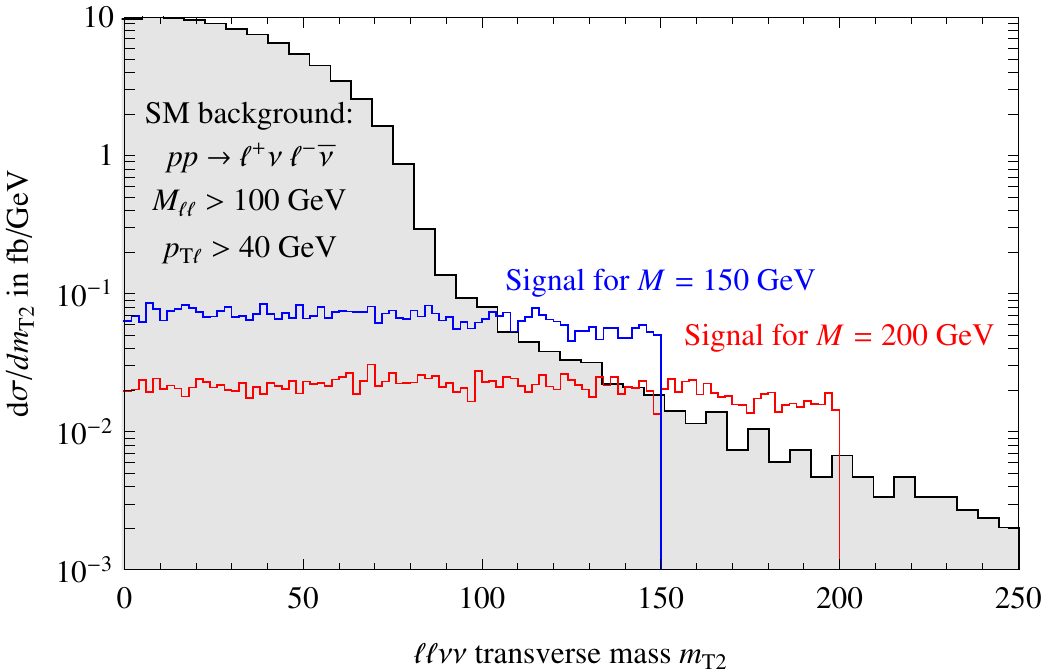}
\caption{\label{fig:4ellbck}\em {\bf scalar di-leptons} signals vs backgrounds: 
$pp\to \mu^+\mu^+\mu^-\mu^-$ (left) and $pp\to \ell^+\nu\ell^-\bar\nu$ (right).}
\end{center}
\end{figure}

\subsection{4 charged leptons}
The $\tilde{E}^2$ singlet couples to right-handed leptons as $\tilde{E}^2 EE$: its signal is
4 charged leptons $\ell^+\ell^+ \ell^-\ell^-$ with
equal invariant mass of the same-sign lepton pairs. 
This signature gets lost if one of the leptons is
a $\tau$  (in such a case the $\mu$ and $e$ spectra from its decays allow in principle to test its polarization), 
and in the worst case where $4\tau$ are produced the signatures are similar to the ones
discussed in~\cite{muonCDF}. 
In the best case one has $\mu^+\mu^+ e^- e^-$ states that violate lepton flavor.
We here focus on the $\mu^+\mu^+\mu^-\mu^-$ case, showing that SM backgrounds 
are well below the total signal cross-section, plotted in fig.\fig{sigma}.
In the SM $\sigma(pp\to \mu^+\mu^-\mu^+\mu^-)=6.8\fb$.
Fig.\fig{4ellbck}a shows, as function of $M_{\rm eff}$, the cross section of such events with the
further requirement that $M_+\equiv M_{\rm eff}(\mu^+, \mu^{+})$ and
$M_-\equiv M_{\rm eff}(\mu^-, \mu^{-})$ differ by less than 25\%.
%\xxx{Roberto, what is the true uncertainty?
%Cuts on the signal still need to be applied (on $p_T$....)}
We see that despite the loose requirement in the difference $|M_--M_+|$ the signal is already clean. Thus at this stage the luminosity for discovery is set by the signal rate only and no tightening of this requirement seems needed.
%We see that the signal is already clean, although its cross section is low.
Furthermore, one can remove the $Z\to \ell^+\ell^-$ background by demanding that opposite-sign
lepton pairs do not reconstruct the $Z$ mass, suppressing the irreducible SM backgrounds by a factor $\sim 10^3$
with respect to fig.\fig{4ellbck}a.
Before concluding that the signal is background-free, one should also consider fake leptons or
backgrounds (such as $\mu$ from $\pi$ decays), that can be suppressed by demanding isolation criteria.
This was achieved by the D0 collaboration, that searched for
similar signals~\cite{D0}, finding that at TeVatron
the $\ell^+\ell^+ \ell^-\ell^-$ signal must have cross section below about 20 fb, which implies
scalars heavier than about $130 \div 150\GeV$~\cite{D0}. As such this kind of signature seems visible at the LHC as soon as the luminosity collected is sufficient to produce an handful of signal events.

\subsection{2 charged leptons and 2 neutrinos}
The $\tilde{E}$ singlet couples to left-handed leptons
with flavour anti-symmetric couplings $\tilde{E} L_iL_j =\tilde{E} (\ell_i \nu_j - \ell_j\nu_i)$
and thereby is somewhat similar to a heavier leptonically-decaying
$W^\pm$, produced with a smaller cross-section.
Thereby the signal of $\tilde{E}$ is more elusive: two opposite-sign leptons accompanied by missing energy.
The signal cross section is the lowest one in fig.\fig{sigma}.

The SM  background process is $pp\to \ell_i \bar{\ell}_j \bar{\nu}_i \nu_j$ with a total cross section of about 2 pb. As this background is mostly due to two-body processes like $pp\to W^+W^-$ and $pp\to ZZ$ it can be reduced requiring final state leptons and missing transverse energy that force the vector bosons to be off-shell. This can be done requiring 
\beq M_{\rm eff}(\ell_i \bar{\ell}_j)> M_Z \label{Zoffshell}\eeq 
and 
\beq {m}_T^2(\ell\nu)\equiv2 E_T^\ell \slashed{E}_T(1-\cos\phi_{\ell\nu}^T)> M_W^2 \label{mT}. \eeq %il BR e' 0.25
%SM backgrounds  have a much larger cross-section $\sigma(pp\to \ell^+ \ell^- \nu\bar\nu) \approx 1.6\pb$ with $\ell=e$
%plus $\mu$ and their main and more difficult component is $\sigma(pp\to W^+W^-)\approx 1.8\pb$
%with leptonically decaying $W$.
%As well known, background events with a single $W^\pm$ can be eliminated 
%by demanding that the  `transverse mass'  of the 
%lepton-neutrino pairs arising from its decays is above $M_W$:
%\beq 
%	{m}_T^2(\ell\nu)\equiv2 E_T^\ell \slashed{E}_T(1-\cos\phi_{\ell\nu}^T)> M_W^2 \label{mT}.
%\eeq %il BR e' 0.25
Here $\slashed{E}_T$ and $E_T^\ell$ are the missing transverse momentum of the neutrino and
of the lepton and $\phi_{\ell\nu}^T$ is the angle between the components of their momenta
transverse to the beam.

In this kind of events $m_T$ cannot be directly computed because there is more than one source of missing transverse energy. However the cut in eq. (\ref{mT}) can be enforced using the variable $m_{T2}$ of~\cite{mT2} and requiring 

\beq m_{T2}^2 \equiv \min \max({m}_T^2(\ell_1 \nu_1),{m}_T^2(\ell_2 \nu_2)) > M_W^2 .\label{mT2cut}\eeq

The cuts of eq.s~(\ref{Zoffshell}), (\ref{mT2cut}) have a mild effect on the signal because leptons and missing transverse energy are generated by an heavy particle which can be on-shell and still pass the cuts. As such, even though the signal rate is not very high, it can result in an excess of event after $\sim$10/fb of luminosity have been collected.

Fig.\fig{4ellbck}b shows an example of the $\tilde{E}$ signal (computed for $M=(150)~200\GeV$ such that $\sigma \approx 12(7)\fb$
times a $60\%$ acceptance) 
%an integrated luminosity of about 10/fb would be enough to observe this signal)
%(computed for 30/fb $M=300\GeV$ such that $\sigma \approx 1.5\fb$) 
compared to the  backgrounds after the requirement $p_T^\ell>40$ GeV and those of eq.s\eq{gencuts}, (\ref{Zoffshell}), (\ref{mT2cut}).
Background tails are even smaller for a lepton-flavor violating signal such as $e^-\mu^+\slashed{E}_T$.

Finally we note that from the distribution in $m_{T2}$ one could in principle measure the mass of $\tilde{E}$ looking at the endpoint of the distribution. 

\section{$2\ell\, 2V$ heavy lepton signals}\label{2l2V}
We consider fermions coupled to a lepton and to the Higgs doublet.
Lepton number is violated if the fermion has a Majorana mass: the two possible
cases are well known as type-I and type-III see-saw; the latter case has LHC manifestations
that have already been studied in~\cite{typeIII}.

We therefore focus on the Dirac case and there are two possibilities\footnote{An early study of the phenomenology of such kind of particles is found in \cite{CCleptons}.}.
\begin{itemize}
\item[i)] the SU(2) triplet
$E_3+\hbox{h.c.}$ coupled as $E_3 LH^*$ with components
 of electric charge $0,1,2$ that can decay as
\beq E^0 \to \nu Z,\qquad
E^+\to \nu W^+ , \ell^+ Z,\qquad
E^{++}\to \ell^+ W^+\eeq
with equal total widths in the $M\gg M_Z$ limit.
Notice that $E^0$ does not decay into $\ell^\pm W^\mp$.
Notice also that although $E^0$ is not a stable DM-like particle,
it looks like that when its decay is invisible, $E^0\to \nu \nu\bar \nu$.
\item[ii)] the SU(2) doublet
$L^{3/2}+\hbox{h.c.}$ coupled as $L^{3/2} \bar EH^*$,
with components
 of electric charge $1,2$ that can decay as
\beq L^+\to \ell^+Z,\qquad
L^{++}\to \ell^+W^+\eeq
with equal widths ($L^+$ does not decay into $\bar\nu W^+$).

\end{itemize}
The primary final states with only charged leptons and heavy vectors
are $\ell^+W^+ ~\ell^-W^-$, $\ell^+Z \ell^- Z$,
$\ell^+ W^- \ell^- Z$. Other similar channels involve neutrinos and higgses.
Lepton flavor can be violated, while lepton number is conserved.
We assume that the leptons $\ell$ produced in heavy-lepton decays are $e$ or $\mu$
rather $\tau$.

\subsection{Production of $pp\to W^+ \to E^{++} E^-$}
This production mechanism has the larger cross-section.
$E^{++}$ decays into $W^+\ell^+$, and we assume that $E^-$ decays into $\ell^- \slashed{E}_T$
giving rise to a $W^+\ell^+~\ell^- \slashed{E}_T$ state, such that this signal exists
for both the heavy  lepton triplet $E$ and doublet $L$.
If the $W$ decays leptonically and  $Z\to \nu\bar\nu$, the final state is $\ell^+\ell^+ \ell^-\slashed{E}_T$
with BR $\approx 4(2)\%$ for the heavy triplet $E$ (doublet $L$). 
If instead $E^-\to Z\ell^-, h \ell^-$ and the $Z$ or $h$ decays hadronically, or if there are jets from QCD initial state radiation,
the signal is 
\beq pp\to \ell^+\ell^+ \ell^-\slashed{E}_T X\eeq  where $X$ denotes extra particles.

In the case the heavy lepton is light enough that a large number of signal events is present,
we can restrict to the cleaner state where $X$ is empty:
%This can be done by vetoing central hard jets, as they are negligibly generated by QCD radiation.
%We study the SM bakgrounds to the resulting  signal:
 \beq
 pp\to \ell^+\ell^+ \ell^-\slashed{E}_T\, . \label{exclusive}\eeq
 
The production of leptons in the SM is mainly given by vector bosons decays. Hence the dominant contribution to the background arise from two-body processes like $pp\to ZW^+$ which can be efficiently suppressed requiring opposite-sign leptons with invariant mass $m_{\rm OS}$
well above $M_Z$, leaving only 3-body backgrounds with smaller cross sections:

\begin{enumerate}
%\item The background is $\sigma(pp\to ZW^+) = 17\pb$, which can be suppressed
%by demanding decays into opposite-sign leptons with invariant mass $m_{\rm OS}$
%well above $M_Z$, leaving only 3-body backgrounds with smaller cross sections.

\item $pp\to \ell^+ \ell^- W^+$ from off-shell $Z$ or $\gamma$ which gives $ \approx 9 (1) \fb$ after demanding $m_{\rm OS}>100 (200) \GeV$
and a leptonic $W^+$ decay.

\item $\sigma(pp\to W^+W^+W^-)\approx 1\fb$ 
% no kinematical cuts on leptons and for $m_h\approx 120\GeV$.
%   We checked that for mh=200 above WW threshold nothing changes
after demanding
that all $W$ decay into $e$ or $\mu$.
\end{enumerate}

The exclusive signal of eq.~(\ref{exclusive})  has no hard jets. However we expect that some hadronic activity will be present due to initial state radiation (ISR). This activity will produce jets that get harder and harder as the hard scale of the process increases, i.e. as the mass of the new particle increases. Thus one should consider additional backgrounds containing jets, although the signal had no jets at the partonic level.  The most relevant background in this class is $pp\to t\bar t W^+$ which has a non negligible cross-section $ \approx 10\fb$ after demanding leptonic $W$ decays. 
%\xxx{check o 3 fb?}
%tt would not have enough isolated leptons. in facts to have three leptons one should require at least one of the b quarks to decay semileptonically and this produces an isolated lepton only in a small fraction of events. Starting from tt->bbWW with 700pb, bb mET l-l+ is 28pb, c l+ b mET l+ l- with an isolated lepton is 280 fb, hence above the ttW->bb3lmET is actually subdominat by a factor 10. However the jet veto will bring this BG at a level below Z/gamma W, therefore the partonic level study is safe.

Showering both the $t\bar{t}W^+$ background and the signal with {\sc Pythia} 8.1 \cite{pythia} we studied the efficiency of a veto on central hard jets finding that $t\bar{t}W^+$ can be  reduced by a factor few times $10^{-2}$ while the signal gets reduced only by a factor $0.3\div0.4$. This allows to neglect this class of backgrounds and renders our study at the partonic level rather reliable.
Processes where leptons are produced by meson decays provide extra backgrounds.
However, the typical lepton arising in such processes is not isolated from the hadronic activity of the event and therefore only ``rare'' events contribute to the background. Assessing the relevance of these kind of backgrounds would require to evaluate the efficiency of lepton isolation cuts, which is beyond our
scope. However Ref. \cite{berger} studied this issue and according to their results seems that the analysis  detailed in the following  should not be spoiled by this source of leptons.

\medskip

\begin{figure}
\begin{center}
\includegraphics[width=0.5\textwidth]{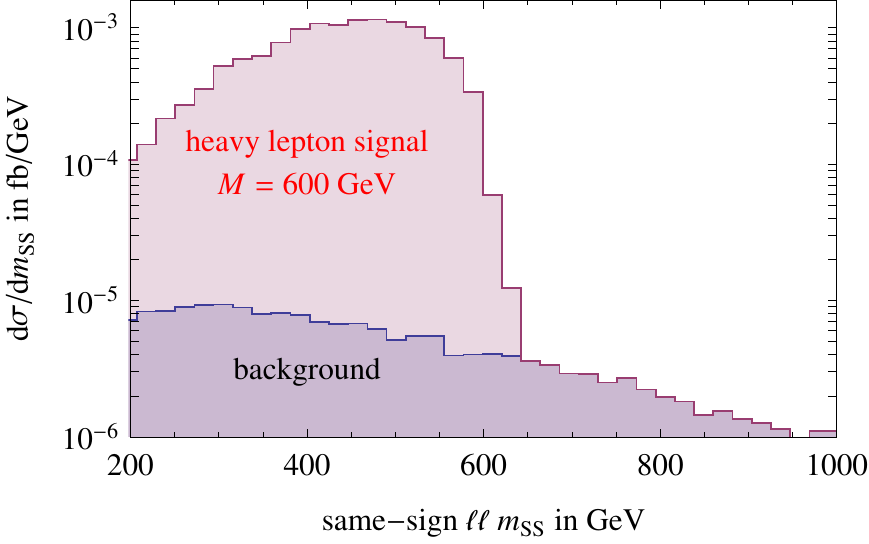}
\caption{\em\label{fig:3lmET} {\bf Heavy lepton} signal+background and background-only distributions of the invariant mass of same-sign leptons for the $pp\to E^{++} \cdot E^-\to W^+\ell^+\cdot \ell^- Z\to \ell^+\ell^+\ell^-\slashed{E}_T$ heavy lepton signal of eq.~(\ref{exclusive}) .}
%We assumed an integrated luminosity of $30/\fb$ and bins of  $22 \GeV$.}
\end{center}
\end{figure}

This signal is characterized by same sign leptons whose invariant mass is a fraction of the mass of the new particle. The mass of the new particle also sets the scale for the missing transverse energy 
$\slashed{E}_T$ and for 
the invariant mass of opposite signs leptons, $m_{OS}$. Therefore we look for the signal in the distribution of the invariant mass of same-sign leptons, $m_{SS}$, in events with large $m_{OS} $ and large $\slashed{E}_T$. 
Figure \ref{fig:3lmET} shows the result for a 
$M=600 \GeV$ 
%900 GeV 
resonance
such that the signal eq.~(\ref{exclusive})  has a cross-section of 
0.5 fb
% 70 ab
%and an integrated luminosity of  
%30/fb 
%100/fb
with the cuts
\beq \slashed{E}_T>200\GeV,\qquad
m_{OS}>150\GeV. \label{mETmOS} \eeq
%\begin{itemize}
%\item $mET>200$ GeV \item $m_{OS}>150$ GeV
%\end{itemize}
% si! i cut sono gli stessi ... cosi' c'e' qualche evento di segnale in piu' ... S/B e' cosi' alto che alla fine i cut contano poco.
The signal extends up to  $m_{\rm SS}< M$ and the signal/background ratio is large, allowing for discovery
as soon as a handful of signal events can be produced, 
after that an integrated luminosity of about 10/fb is collected.
%100/fb

For higher masses $M$ the signal cross section of eq.~(\ref{exclusive}) drops quickly. Hence it would be interesting to study the less clean final states where $X$ is non empty. In this case the evaluation of backgrounds where jets appear from ISR and leptons arise from meson decays would be necessary. 
We do not perform this study, however Ref. \cite{berger} examined some similar case in the context of supersymmetric signatures like eq. (\ref{exclusive}) and found that a cut in missing transverse energy like eq.(\ref{mETmOS}) provides good rejection of this kind of background where leptons are not generated together with $ \slashed{E}_T $ .

%In this case backgrounds like $pp\to t\bar{t} \to b\ell^+~ \bar c \ell^+ \ell^-~\slashed{E}_T$  becomes a potentially problematic background
%which needs to be explored.

Production of $pp\to W^+ \to E^{+} E^0$ has the same rate of the production of $E^{++} E^-$ studied above in this section,  and its experimental signatures are a subset of the ones already considered in the type-III see-saw context~\cite{typeIII}.

%In the present context however the features of the missing ener

%However in the the present context it is particularly delicate the case where the $E^0$ get undetected or decays invisibly. In facts in this kind of events there is a large missing transverse energy which in principle could arise from any number of invisible particles. Assessing that this large missing transverse energy is originated by one single body escaping the detector is a check of the hypothesis embodied in eq. (\ref{eq:Z2odd}).

%\xxx{This check can be done studying the invariant transverse mass of the system recoiling against the visible particles. This should have a peak and an end-point at the value of the mass of $E^0$}.

\begin{figure}
\begin{center}
\includegraphics[width=0.45\textwidth]{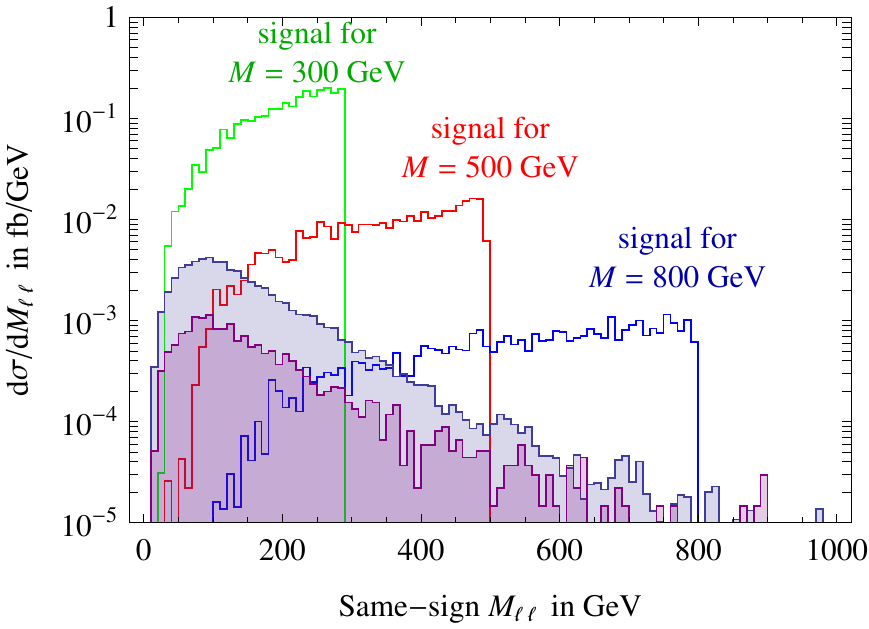}\qquad
\includegraphics[width=0.45\textwidth]{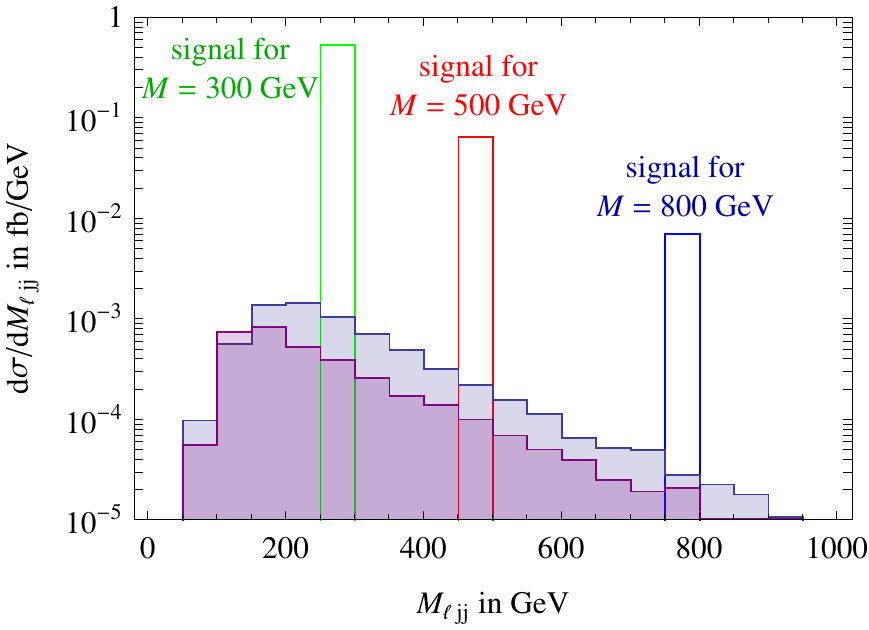}
\caption{\label{fig:2l2V}\em {\bf Heavy lepton} $pp\to E^{++}E^{--}\to \ell^+W^+\cdot \ell^-W^-\to 
\ell^+\ell^+\slashed{E}_T\cdot \ell^-jj$   signal 
for $M=\{300,500,800\}\GeV$
vs $t\bar t W^+$ (upper) and $ZW^+W^-$ (lower) backgrounds.}
\end{center}
\end{figure}

\subsection{Production of $pp\to Z,\gamma \to E^{++} E^{--}$}
The signal is $\ell^+ W^+~\ell^-W^-$ with equal invariant masses for the first and second pair.
The possible final states are:
\begin{itemize}
\item The $\ell^+\ell^- 4j$ signal has the highest BR $\approx 46\%$ but also
a huge background of about $10\pb$ if the two leptons have the same flavor.
Given that the full event can be reconstructed
various cuts and selections on invariant masses can be performed.

\item The $\ell^+\ell^-\ell^+\ell^- \slashed{E}_T$ signal has the smallest BR $\approx 4.5\%$.
The backgrounds are:
 $\sigma(pp\to ZWW)\approx 200\fb$ reduced down to $0.2\fb$ after restricting to leptonic decays, 
 $\sigma(pp\to ZZ)\approx 10\pb$ reduced down to $1\fb$ after restricting to $Z \to\tau^+\tau^-$
and leptonic $\tau$ decays, 
$\sigma(pp\to t\bar{t}) \approx 700\pb$ reduced by a large amount by leptonic BR and 
jet veto and lepton isolation cuts.
These backgrounds can be further suppressed demanding that the invariant mass of
opposite sign leptons is above $M_Z$, or if the signal violates lepton flavor.

\item
The $\ell^\mp\ell^\mp\ell^\pm\slashed{E}_T 2j$ signals have BR $\approx 14.4\%$ each.
The main background is $\sigma(pp\to t \bar t W)\approx 320\fb$, that becomes $\approx 3\fb$ after imposing the appropriate decay modes; it can be suppressed exploiting the kinematical features of the signal (large invariant masses of same-sign leptons, $M_\mathrm{eff}(jj) = M_W$).
The $\sigma(pp\to ZW^+W^-)\approx 100\fb$ background can be eliminated imposing 
$M_{\ell^+\ell^-}>100\GeV$. With this cut one is left with the $pp\to \ell^+\ell^-W^+W^-$ background;
its cross section is
 $\sigma\approx 0.4\fb~(0.06\fb)$ before (after) imposing the appropriate $W$ decays.

Another background is $pp\to t\bar{t}$, with $t\to b W \to c WW$ followed by leptonic $W$ decays;
devising cuts that reduce its large cross section by a large enough factor is mainly an
experimental issue that we do not address.

This signal allows to measure the mass of the heavy lepton in two different ways:
as the endpoint of the same-sign leptons invariant mass distribution (fig.~\ref{fig:4ellbck}a);
and as a peak in the invariant mass of the two jets with the opposite-sign lepton (fig.~\ref{fig:4ellbck}b).
In both variables the signal is well above the backgrounds.
\end{itemize}

%At this level, the SM bakgrounds (summed over $\ell=\{e,\mu\}$)
%are $\sigma(pp\to\ell^+W^+ ~\ell^-W^-)\approx 5\fb$,
%$\sigma(pp\to\ell^+Z ~\ell^-Z)\approx 1.7\fb$
%with $M_{\ell^+ \ell^-}$ peaked on $M_Z$ or on 0?

%

%The two best signatures seem to be:
%\begin{itemize}
%\item[a)] ($BR\sim1/3$)
%$(\ell^+ jj)(\ell^- jj)$ with equal invariant mass of the two groups of particles
%and with jet pairs that reconstruct a $W$.

%BCK1: calcolare $\sigma(pp\to \ell^+ \ell^- 4j)$ with $M_{\ell^+\ell^-} > M_Z$
%BCK2: $pp\to W^+ W^- 4j$ con $W\to \ell$.

%\item[c)] ($BR\sim1/3$)
%$(\ell^+ jj)(\ell^- \ell^-\bar\nu)$ 

%\item[b)] $(\ell^+ (Z\to \ell^+\ell^-))(\ell^- (Z\to \ell^+\ell^-))$.
%\end{itemize}

\section{$2j\,2V$ heavy-quark signals}\label{2q2V}
The new particles are SU(3) triplets, that decay into a quark and a $W$ or $Z$ or $h$.
The flavor of the quark is unknown, and again we will consider a generic light quark
rather than a $b$ or a $t$.

If both $W$s decay hadronically, it seems feasible with appropriate cuts to identify the
resulting $6j$ signal in the QCD multi-jet background.
If instead one $W$ decays leptonically and the other hadronically, the resulting
$4j~\ell\slashed{E}_T$ signature (with invariant masses $M_{jj}=M_W$ and $M_{jW}=M$)
emerges over the $t\bar{t}+$ jets and $W+$ jets backgrounds,
as previously studied for the analogous 4th-generation signal in~ \cite{WWqq}.
In such a scenario the new particle, called $t'$, decays as $t'\to W b $, while
$t'\to Zu$ decays are not present at tree level, as the $t'$ gets its mass from
the same Yukawa couplings that give rise to its decay, so that $Z$ couplings are flavor conserving.

\begin{figure}[t]
\begin{center}
\includegraphics[width=0.5\textwidth]{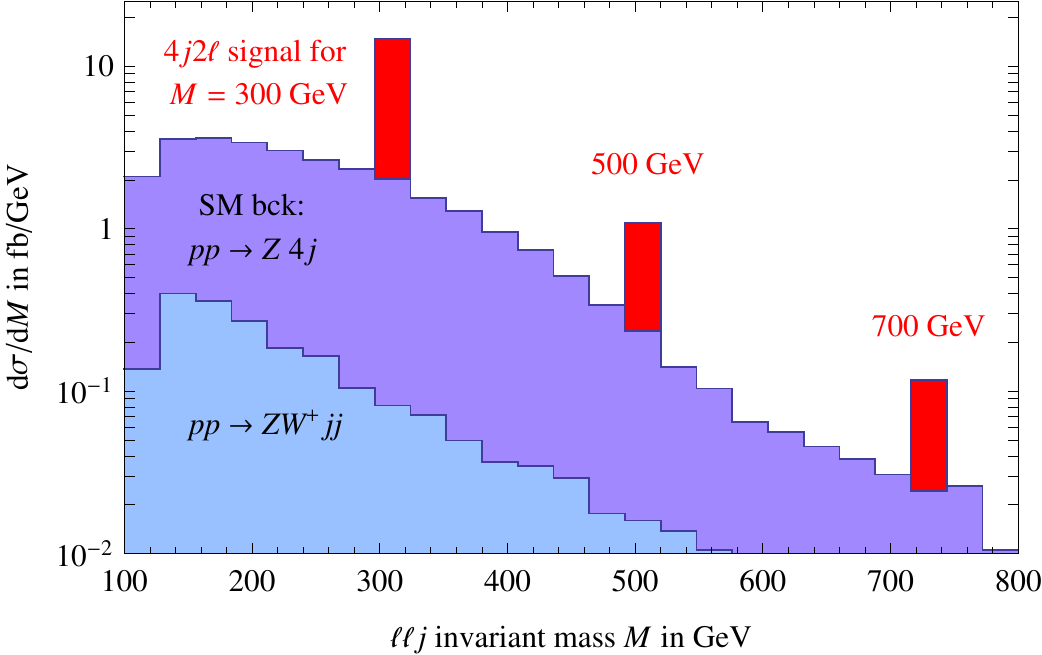}
\caption{\label{fig:2q2V}\em {\bf Heavy quark} $pp\to jjWZ\to 4j\,2\ell $ signal vs SM backgrounds.}
\end{center}
\end{figure}

In order to avoid problems with precision data and flavor,
we instead consider new non-chiral heavy quarks, that therefore have
a mass term invariant under $\SU(2)_L\otimes{\rm U}(1)_Y$ as well as small Yukawa couplings:
decays into $Zq$ and $hq$ are necessarily present, with a branching ratio of about $25\%$ each.
Ignoring the still unknown Higgs boson, we obtain in our scenario the cleaner $jjWZ$ and $jjZZ$ events.
The best signature is $4j\,2\ell$,
produced in about $1\%$ of the signal events, when $W,Z\to jj$ and $Z \to \ell^+\ell^-$
with $\ell=\{e,\mu\}$.
The main SM backgrounds are:
\begin{enumerate}
\item $\sigma(pp\to4j\,Z \to 4j2\ell) \approx 10 \pb$.
\item $\sigma(pp\to jj WZ\to 4j2\ell) \approx 0.4\pb$.
\item $\sigma(pp\to t\bar{t}jj \to 4j2\ell) \approx 10 \pb$ and $ \sigma(pp\to4j2W\to 4j2\ell) \sim 0.2\pb$
become subdominant with respect to the previous backgrounds after imposing $M_{\ell\ell}=M_Z$.
\end{enumerate}
We select the events imposing the standard isolation and detection cuts of eq.\eq{gencuts}
(the acceptance is $\approx 20\%$ for the signal) and
that some combination of jets satisfies
$M_{jj}=M_W\pm 10\GeV$ and $M_{\ell\ell j} = M_{jW}$ within 10\%.
In fig.\fig{2q2V} we plot the background~\cite{AlpGen} and the signal as function of this latter invariant mass variable,
such that the signal is a peak at the heavy-quark mass. The signal events are generated for the case of an $\SU(2)_L$ singlet quark. Bigger $\SU(2)_L$ multiplet will result in bigger cross sections due to the possibility to produce further weak-isospin replicas.

\medskip

Table~\ref{tab:listf} lists 7 different `heavy quarks' (3 SM-like and 4 with exotic quantum numbers),
so that one wonders which LHC observable could discriminate among them.
A partial discrimination can be done measuring their mass and their production cross section: as already pointed out
heavy quarks can be singlet, doublet or triplet under $\SU(2)_L$, and one doublet (triplet) has
a production cross section 2 (3) times larger than one singlet.
Furthermore heavy quarks with exotic charges can only decay into $W^\pm q$, so that their presence leads to
an enhanced ratio between $jjWW \div jjWZ\div jjZZ$ signals.
Discriminating $W$ from $Z$ presumably can only be done observing the signals from their leptonic decays.
Observing all the three signals above would therefore allow to probe
the presence of exotic heavy quarks.
The presence of top or bottom quarks (which give rise to extra specific signatures)
would provide an extra handle.

\section{$2\ell \,2q$ lepto-quark signals}\label{2l2q}
New scalars  that couple to a lepton and a quark (and thereby commonly called Lepto-Quarks, LQ)
give rise to final states containing 2 leptons and 2 quarks.
The resulting LHC signals and capabilities have been already extensively explored~\cite{LQ},
so that we will not focus on such topic.

We notice that according to our classification some particles both give rise to LQ signals together with the QQ signals
(as well as possible  violation of baryon-number) discussed in  section~\ref{4q}.

\section{Conclusions}\label{con}
We classified all new scalar or fermion multiplets of the SM gauge group
such that one new particle can couple to a pair of SM particles (either the Higgs boson or quarks or leptons)
in a gauge and Lorentz invariant way.
Such `Minimal Matter' scenario is orthogonal to the standard ideology, that instead assumes a sector of new particles
supplemented with an ad-hoc discrete symmetry that forbids all the above couplings
such that new particles can only couple in pairs to the SM sector and
the lightest new particle can be a stable Dark Matter candidate.
However this symmetry is not necessary; 
automatically stable `Minimal' Dark Matter candidates were proposed in~\cite{MDM}.

%, finding
%a few multiplets (mainly a Majorana weak 5-plet) that
%cannot have any such coupling  and contain a neutral component,
%thereby giving automatically stable Dark Matter candidates.

Assuming that the new couplings are small avoids all current {\em indirect} constraints 
leaving {\em direct} production at LHC as the main probe: 
indeed new particles are produced in pairs via their gauge interactions,
so that we computed their production rate (fig.\fig{sigma}),
while the smallness of their couplings to SM particles only adds a small decay width.
We obtained a set of well defined new-physics signatures and presented the dedicated searches
that can allow to see such new physics among the huge backgrounds present at LHC.
These signatures fall into five main classes:
\begin{enumerate}
\item The well known {\em scalar lepto-quarks}, that couple to leptons and quarks.
\item {\em Scalar di-quarks} couple to two quarks and
lead to $4j$ signatures with peaks in the two $jj$ invariant masses; we
explored how they could be discovered as a small excess as well as isolated
as a clean peak (fig.\fig{4j}).

\item {\em Scalar di-leptons} couple to two leptons and
lead to $4\ell$ signatures: either $\ell^+\ell^-\ell^+\ell^-$ (which
easily emerges over the SM backgrounds, fig.\fig{4ellbck}a) or $\ell^+\ell^- \slashed{E}_T$ (which emerges over
the SM backgrounds after considering a dedicated transverse mass $m_{T2}$ of the system, fig.\fig{4ellbck}b).

\item {\em Heavy leptons} couple to leptons and to the Higgs boson and
lead to $\ell \ell VV$ states, where $V$ is either a $W$ or a $Z$ or a higgs,
giving rise to various signatures: we exemplified the observability of the
$\ell^+\ell^+\ell^-\slashed{E}_T$ (fig.\fig{3lmET}) and $\ell^+\ell^+ \ell^-\slashed{E}_T jj$
(fig.\fig{2l2V}) signals.

\item {\em Heavy quarks} couple to quarks and to the Higgs boson
leading to $jjVV$ states, that are best seen as $4j\ell\slashed{E}_T$ or
$4j2\ell$ (fig.\fig{2q2V}) final states.
Their ratio indicates the $W$ fraction in $V$: a high $W$ fraction indicates the presence of
heavy quarks with exotic electric charges 4/3 or 5/3, that can therefore decay only into $Wq$.

\end{enumerate}
A few multiplets (such as the scalar triplets of type-II see-saw)
can have couplings of different types, violating the conserved
global symmetries of the SM and leading to special signatures~\cite{typeII,Raidal}.

Final states involving no missing energy $\slashed{E}_T$ and the lack of edges
in observable related to $\slashed{E}_T$ allow to discriminate this scenario from
the standard scenarios where a DM LSP-like particle carries away $\slashed{E}_T$.

\paragraph{Acknowledgements} 
This work is supported by the Swiss National Science Foundation under contract No. 200021-116372. We thank  Fulvio Piccinini for help in using weighted events in {\sc AlpGen}. We thank the authors of {\sc FeynRules} for support in the implementation of our models. We thank  Ian Lewis for remarking a mistake in Table \ref{tab:lists}.

\appendix

%\section{(Detector)}

%

%The $e$ (or $\mu$) energy is measured as [according to TaoHan's reading of the CMS TDR]
%$$\frac{\Delta E}{E} = 0.55\% \oplus \frac{5\%}{\sqrt{E/\GeV}}$$
%and the $\mu$ tracking measures
%$$\frac{\Delta p_T}{p_T}=15\% \frac{p_T}{\TeV} \oplus \frac{0.5\%}{\sqrt{\sin\theta}}$$
%and $\oplus$ denotes uncertainties to be summed in quadraure.
%\xxx{Non mi \`e ancora ben chiaro cosa significhi e quale sia l'errore su $M_{\rm eff}$}

%Charge lepton misidentification is at the few $\%$ level.
%(Tops can fake SS $\ell$)

%
%For a jet
%$$\frac{\Delta E}{E} = 3\%\oplus \frac{50\%}{\sqrt{E/\GeV}}$$

\def\baselinestretch{1.05}

\newpage

\label{add1}

\normalsize
\section*{Addendum: Minimal Matter at the LHC at 7 TeV}
\setcounter{equation}{0}
\renewcommand{\theequation}{\arabic{equation}B}

LHC is running at reduced $\sqrt{s}=7\TeV$ with the hope of accumulating an integrated luminosity of at least 1/fb by the end of 2012. Compared to the design $\sqrt{s}=14 \TeV$ considered in the main text, the partonic luminosities are reduced by a factor 2 to 10 for a partonic center of mass energy between $100 \GeV$ and $1 \TeV$, however there is still a significant increase compared the Tevatron. Therefore LHC is expected to superseed Tevatron to make discoveries and in placing bounds even in this ``preliminary'' run. In the following we discuss the signals of minimal matter that can be searched for in the ongoing LHC run.

\begin{figure}[t]
$$\includegraphics[width=0.45\textwidth,height=9cm]{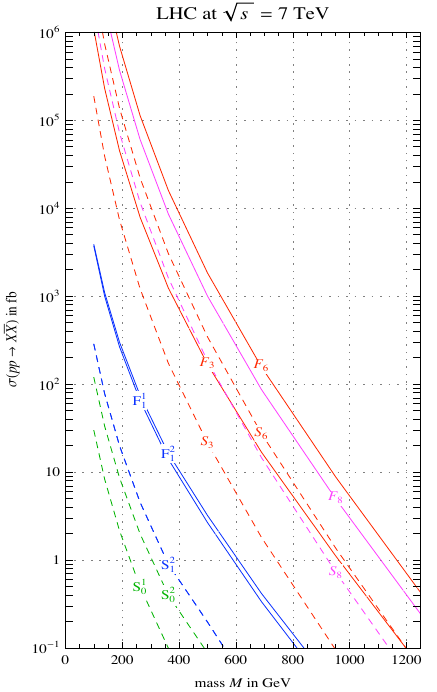}\qquad
\includegraphics[width=0.45\textwidth,height=9cm]{sigmaTeVatron}$$
\caption{\label{lhc7} \em Cross sections for pair production via gauge interactions at leading order
in hadronic collisions
of new particles $X\bar X$ labelled as
$P^Q_{T_3}$ where $Q$ is the electric charge,
$T_3$ is weak isospin and
$P=F~(S)$ for a fermion (scalar).
Cross sections of colored particles (in red) negligibly depend on their electroweak interactions,
so we adopted the simplified notation $P_d$ where $d=\{3,6,8\}$
for color triplets, sextets and real color octets.
Couplings and MSTW 2008 pdf are renormalized at $Q^2=\hat{s}$. }
\end{figure}

The total leading-order
cross-sections for the LHC running at $7 \TeV$ are given in figure \ref{lhc7} with the same notation of figure \ref{fig:sigma}. 
%penso sia $\sigma(M,s) \propto f(s/M^2)/s^2$
Neglecting the logarithmic variation of parton distributions with energy we have
$\sigma(M,s) \propto f(s/M^2)/s$,  and consequently the following scaling:
$\sigma(M,s) \approx \sigma(M/2, s/4)/4$.
This roughly means that reducing the energy of LHC $\sqrt{s}$ by a factor of $2$, reduces its reach in mass by a factor 2.
In a precise analysis, one needs to take into account also the factor of 4 
(less important than the stronger variation of the function $f$), the luminosity of the collider,
and the variation in the background rate.

%  slava arguments of 1/4 rescaling ... exact but accademic, no information about the reach can be obtained from that argument 

The on-going run of LHC aims at collecting 1/fb of integrated luminosity: therefore it will be sensitive to processes with cross-section larger that 10-100 fb. Using the cross sections of figure \ref{lhc7}, one can see the reach in the mass of the new particles. 
For a more detailed analysis, we selected 4 representative cases, one for each type of signals discussed in this paper.
Namely we will discuss: % \begin{enumerate}
i) one diquark: the scalar color octet $S_8$ ($4j$ signal);
ii) one heavy quark: the fermionic color triplet $F_3$ ($2j\,2V$ signals);
iii) one dilepton: the scalar weak singlet of hypercharge 2 called $S_0^2$ ($4\ell$ signals);
iv) one heavy lepton: the weak triplet of hypercharge 2 called $F_1^2$ ($2\ell\,2V$ signals).
%\end{enumerate}

\begin{figure}
$$\includegraphics[width=0.45\textwidth]{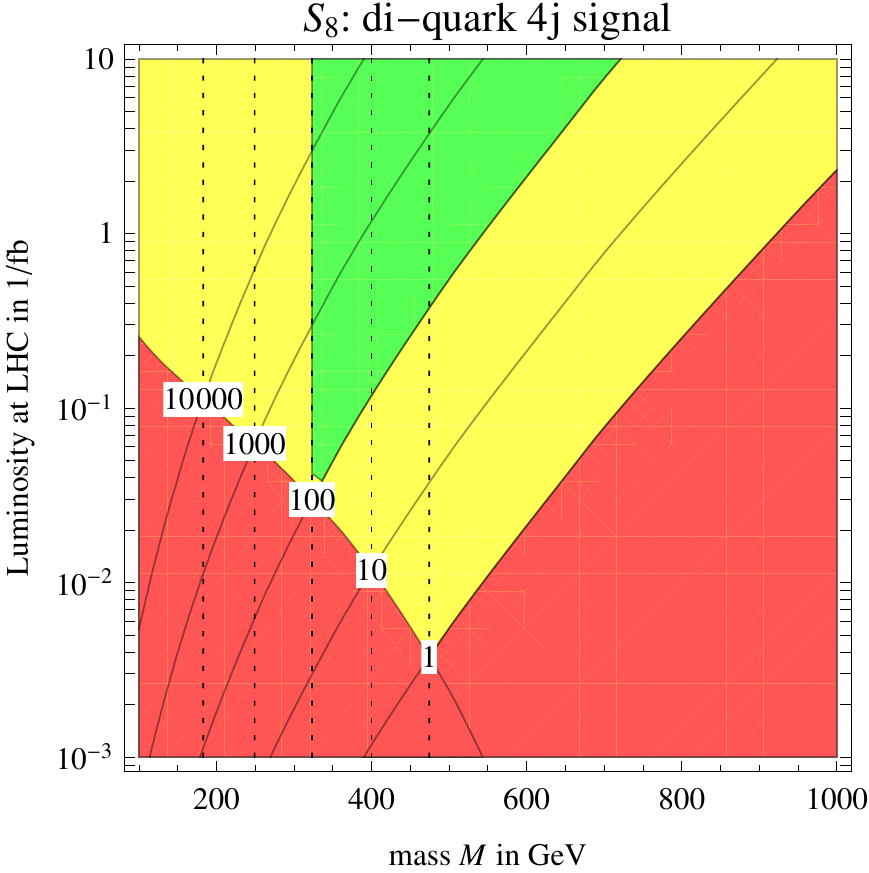}\qquad
\includegraphics[width=0.45\textwidth]{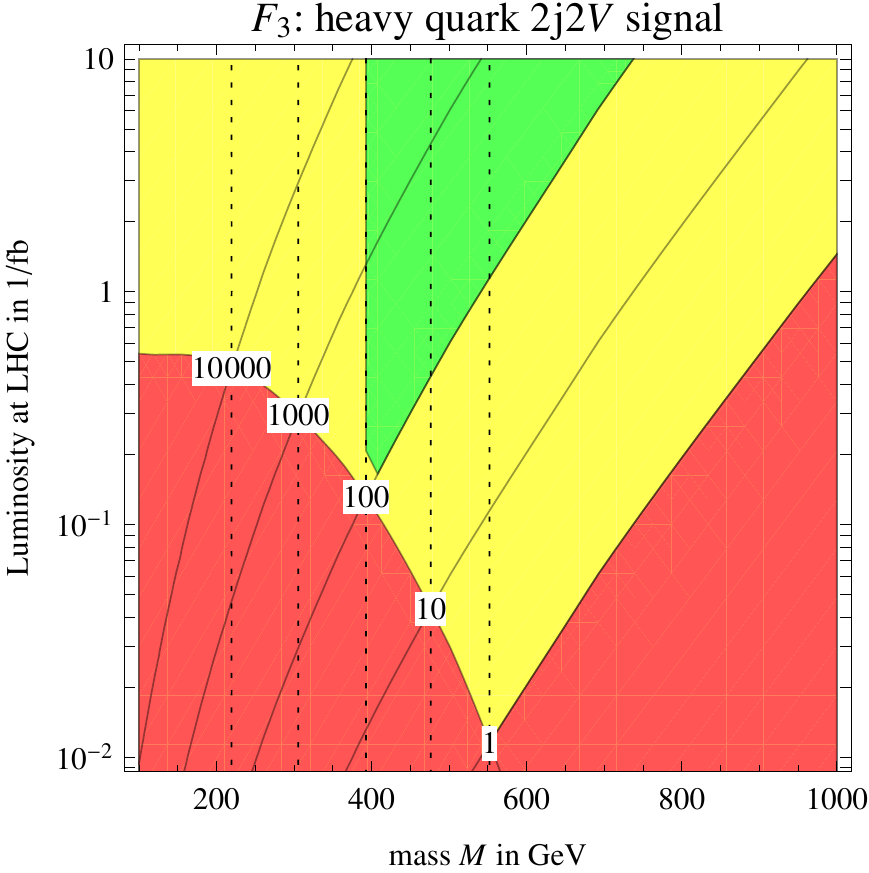}$$
$$\includegraphics[width=0.45\textwidth]{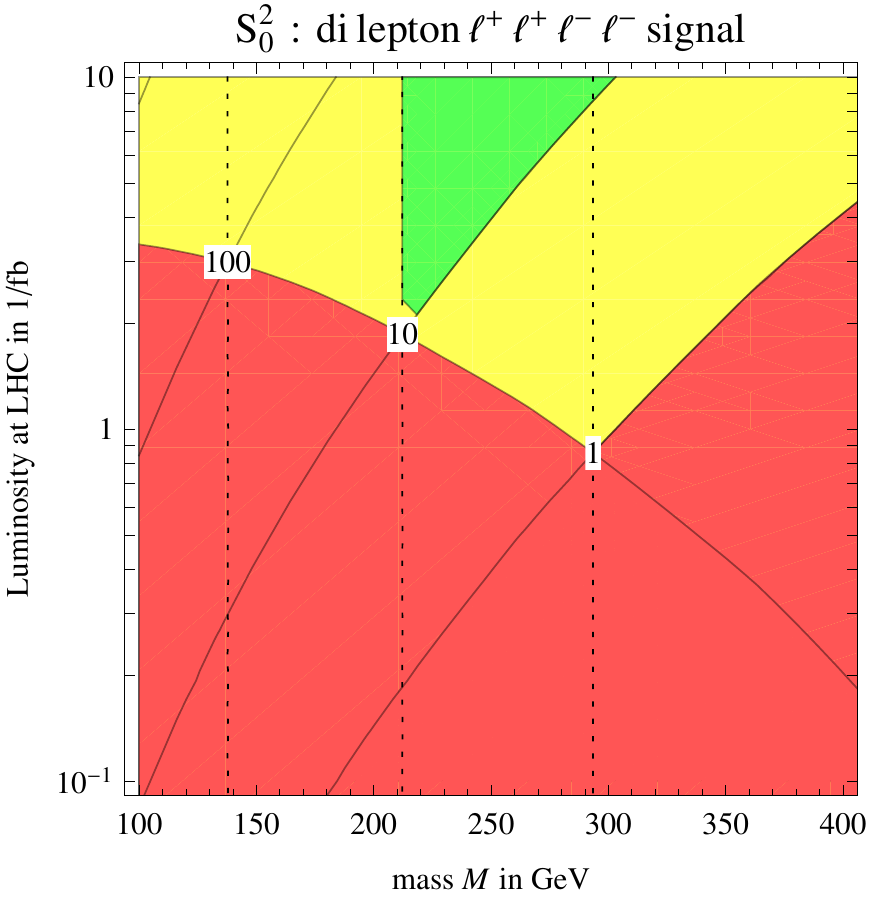}\qquad
 \includegraphics[width=0.45\textwidth]{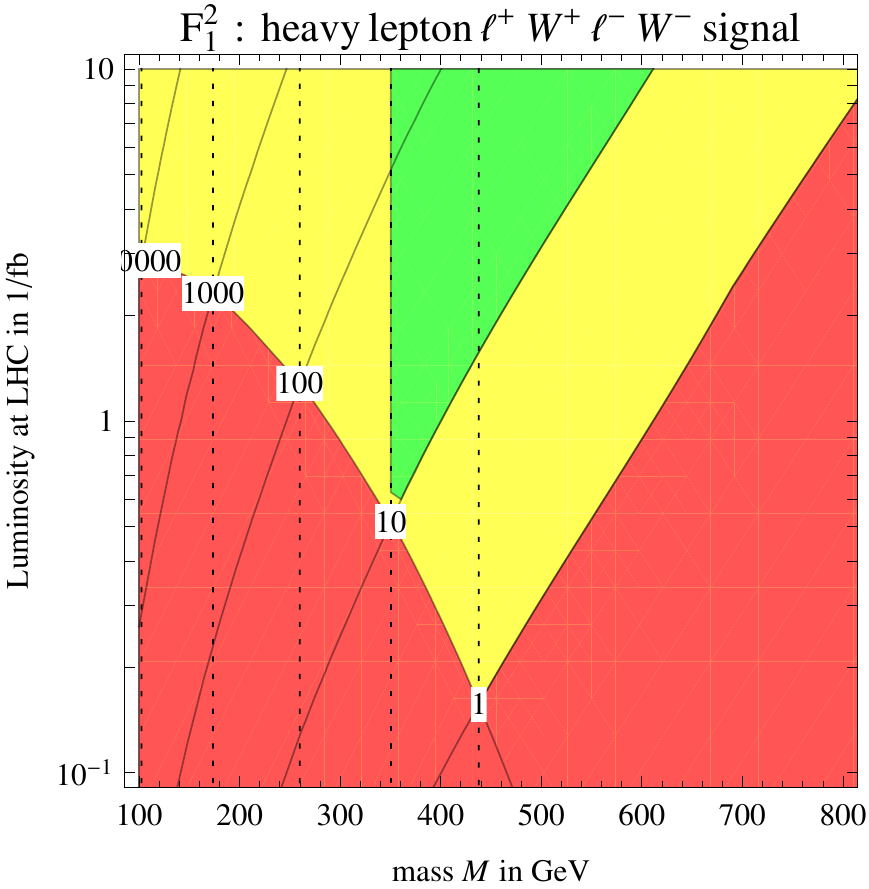}$$
\caption{\label{fig:lhc7reach} \em The reach of LHC running at $\sqrt{s}=7 \TeV$  for the discovery of representative examples of minimal matter candidates. The upper row is for the colored particles $S_8$ (left) and $F_3$ (right). The lower row is for non-colored particles $S_0^2$ (left) and $F_1^2$ (right). The continuous (dahsed) contrours show the number of signal events at LHC (TeVatron).
In all the panels the green area is the regions where LHC has enough luminosity for a discovery that is not accessible to TeVatron. The red regions are those where LHC is either not competitive with TeVatron or has less than 1 event. In the low mass part of the yellow region both LHC and TeVatron can go for a discovery, while in the high mass part of the yellow region both TeVatron and LHC lack events, but LHC has more.}
\end{figure}

Our results for the four relevant signatures at the LHC running at 7 TeV are shown in figure  \ref{fig:lhc7reach}.
%In figure  \ref{fig:lhc7reach} we study the capabilities of LHC at 7 TeV to the 
%four signals we propose: focusing on the following representative Minimal Matter particles:
%the di-quark $S_8$;
%the heavy quark $F_3$;
%the heavy lepton $F_1^2$;
%the di-lepton $S_0^2$.
The relevant variables are the particle mass $M$ on the horizontal axis
and the collected integrated luminosity at LHC on the vertical axis.
When comparing with  TeVatron, we fix  its integrated luminosity at 10/fb, a realistic value.
To assess the reach we estimate that, depending on how clean the signature is, a minimum number of events, $N_{\rm disc}$, must be produced for the discovery. For the cases of $S_0^2$ and $F_3$ we also show current bounds from TeVatron \cite{D0,F3}. From the same bounds we estimate that, once the backgrounds are taken into account, the number of signal event needed for a discovery are about 10 for $S_0^2$ and about 500 for $F_3$. For $F_1^2$ we do not have found a specific search and we assume that 10 events are needed for the discovery. For the case of di-jet resonances the results of section \ref{4q} allow us to estimate that order 500 event should be enough for a discovery.

The coloured areas in figure \ref{fig:lhc7reach} have the following meaning:
\begin{itemize}\color{red}
\item \color{black}In the red area, no discovery can be done at the LHC, either because the number of signal events at LHC is less than one (at high mass), or because TeVatron would have more events (at low mass).\color{yellow}
\item \color{black} In the yellow region, a discovery at LHC is unlikely. More precisely, in
 the low mass part of the yellow region both LHC and TeVatron would have more than $N_{\rm disc}$ events. Therefore this part of the yellow region represents the cases where the experiments on the two accelerators can compete for a discovery.  
In the high mass part of the yellow region  LHC can see more events than TeVatron, but less than $N_{\rm crit}$, as such in this region we expect the discovery to not be possible due to lack of statistics. \color{green}
\item \color{black} A discovery at LHC is expected in the green region. Indeed LHC experiments would collect
 more than $N_{\rm disc}$ events, while TeVatron experiments would have less than $N_{\rm disc}$ events and therefore not enough statistics for discovery. 
\end{itemize}
We see that, for colored particles, LHC starts to explore new regions of parameter space once $\sim$100/pb of integrated luminosity is reached.
For uncolored minimal matter particles, LHC starts to superseed  TeVatron once  a luminosity of $\sim $1/fb is accumulated.
 Fig.~\ref{fig:4jlhc7} exemplifies how a di-quark signal would appear.

\end{document}